\documentclass[10pt]{iopart} 

\usepackage{iopams}  
\usepackage{graphicx}
\usepackage{subcaption}
\usepackage{wrapfig} 
\usepackage{xcolor}
\usepackage{hyperref}
\usepackage{cite}
\hypersetup{
  colorlinks=true,
  linkcolor=purple,
  citecolor=purple,
  urlcolor=purple,
  linktoc=all,
}
\makeatletter
\renewcommand\subsubsection{\@startsection{subsubsection}{3}{\z@}%
  {1.2ex plus 0.3ex minus 0.2ex}
  {0.3ex}
  {\normalfont\itshape}}
\makeatother
\newcommand{\ExB}{\mathbf{E}\times\mathbf{B}}
\begin{document}
\title[Effects of Tungsten Radiative Cooling on DIII-D plasmas]{Effects of Tungsten Radiative Cooling on Impurity, Heat and Momentum Transport in DIII-D Plasmas }
\vspace{-10pt}
\author{A.~Tema~Biwole$^{1}$, T.~Odstr\v{c}il$^{2}$, X.~Litaudon$^{3}$, 
S.~Shi$^{4}$, D.~Ernst$^{1}$, C.~F.~B.~Zimmermann$^{5}$, J.~Lestz$^{2}$, 
N.~T.~Howard$^{1}$, P.~Rodriguez-Fernandez$^{1}$, F.~Khabanov$^{6}$,
F.~Turco$^{5}$, C.~Perks$^{1}$, P.~Manas$^{3}$, D.~Fajardo$^{7}$, 
S.~K.~Kim$^{9}$, L.~Schmitz$^{8}$, H.~Wang$^{2}$, W.~Boyes$^{5}$, 
S.~Ding$^{2}$, B.~Victor$^{10}$, C.~Christal$^{2}$, C.~Lasnier$^{10}$, 
T.~M.~Wilks$^{1}$, G.~McKee$^{6}$}
\vspace{2pt}

\address{$^{1}$ MIT Plasma Science and Fusion Center, Cambridge, MA 02139, USA}
\address{$^{2}$ General Atomics, San Diego, CA 92121, USA}
\address{$^{3}$ CEA, IRFM, F-13108 St-Paul-Lez-Durance, France}
\address{$^{4}$ Oak Ridge Associated Universities, Oak Ridge, TN, USA}
\address{$^{5}$ Department of Applied Physics and Applied Mathematics, Columbia University, New York, NY 10027, USA}
\address{$^{6}$ University of Wisconsin--Madison, Madison, WI 53706, USA}
\address{$^{7}$ Max-Planck-Institut f{\"u}r Plasmaphysik, D-85748 Garching, Germany}
\address{$^{8}$ University of California, Los Angeles, Los Angeles, CA 90024, USA}
\address{$^{9}$ Princeton Plasma Physics Laboratory, Princeton, NJ 08543, USA}
\address{$^{10}$ Lawrence Livermore National Laboratory, Livermore, CA 94551, USA}

\vspace{2pt}

\ead{\textcolor{blue}{biwole@mit.edu}}
\vspace{-0.95pc}
\begin{abstract} A first-of-its-kind experiment was conducted in the DIII-D tokamak under WEST similarity constraints on plasma shape and core parameters. This work presents a detailed transport study comparing a reference regime dominated by intrinsic carbon radiation and a high-radiation regime resulting from controlled tungsten (W) injection using the Laser Blow-Off system, with a core tungsten concentration $n_{\mathrm{W}}/n_e \sim 3\times 10^{-4}$  and a radiated-power fraction $f_\mathrm{rad}>0.5$. The W-induced radiative cooling lowered the electron temperature, thereby decreasing $T_e/T_i$  and stabilizing trapped-electron-mode (TEM) turbulence. This transition in turbulence regime reduced momentum and ion thermal diffusivities, yielding ion temperature peaking and a factor-of-two increase in toroidal rotation. At the outer plasma region, enhanced E×B shear and increased collisionality further suppressed ion-scale turbulence, causing a sharp drop in ion heat flux. Consequently, impurity transport, predominantly turbulent in the low-radiation regime, acquired a strong neoclassical inward W convection during radiative cooling, bootstrapping the cooling cycle. Despite $f_\mathrm{rad}>0.5$, radiative collapse was not observed, likely owing to collisional ion-to-electron energy exchange acting as an electron-energy reservoir, together with $1/1$ MHD activity modulating the radiated power through core impurity neoclassical $T_i$-screening. These results support preparation for a tungsten wall change in DIII-D by elucidating tungsten-induced turbulence stabilization. They also provide key insights for interpreting plasma performance in WEST and are relevant to future reactors expected to operate with radiating tungsten-walled plasmas.
\end{abstract}
\vspace{-1pc}

\submitto{\NF} 

\maketitle
\clearpage
\ioptwocol
\setcounter{tocdepth}{3}   
\tableofcontents
\markboth
{Effects of Tungsten Radiative Cooling on Transport in DIII-D}
{A. Tema Biwole \textit{et al.}}
%
\section{Introduction}
The adoption of tungsten (W) as the primary first-wall material in ITER \cite{Loarte_2025}, SPARC \cite{Rodriguez_Fernandez_2022,Creely_2023}, future Fusion Pilot Plants (FPPs), and several emerging commercial reactor concepts \cite{Lion_2025,Guttenfelder_2025} continues to fuel strong interest in understanding how high-Z impurities influence core confinement and the global power balance of high-performance plasmas. Tungsten combines excellent thermal resilience with low sputtering yields and low tritium retention compared to carbon at reactor-relevant temperatures, yet even trace core concentrations can cause substantial radiative cooling. For ITER, modeling predicts that core W concentrations of the order $\sim 10^{-4}$ can lead to unacceptable radiative losses \cite{P_tterich_2013,Fajardo_2024}. 

In plasmas where impurity-induced radiative cooling does not trigger a disruption, it can still significantly alter the underlying heat, momentum, and impurity transport, making an understanding of W cooling essential for defining the operational window of future fusion reactors. Several present-day tokamaks already operate with full or partial W walls, providing experimental environments in which high-Z impurity transport can be studied under a range of conditions relevant to reactor design. 

Devices such as ASDEX Upgrade \cite{Herrmann_2003,NEU2003367,Neu_2009}, JET \cite{Matthews_2011}, EAST \cite{Luo_2017}, KSTAR \cite{Ko_2024} and WEST \cite{Bucalossi_2024,Bourdelle_2015,Richou_2015} collectively contribute important insight into tungsten sourcing, screening, accumulation, and radiative balance in magnetically confined plasmas. DIII-D \cite{Holcomb_2024}, while operating with a carbon wall, provides unique leverage for impurity studies: tungsten can be injected in a controlled and well-quantified manner, and the machine’s exceptional diagnostic suite enables detailed characterization of W cooling and impurity transport under reactor-relevant conditions. Complementarity between DIII-D and WEST has motivated the launch of a new coordinated similarity program operating in an ITER-relevant, hybrid-like regime. 

Hybrid-like plasmas are characterized by stationary, high-performance operation with peaked pressure and current profiles and a central safety-factor profile close to unity ($q_{\mathrm{min}} \sim 1$--$1.2$), while avoiding large sawtooth activity. This behavior is typically sustained by anomalous current diffusion mediated by benign core MHD activity, enabling improved confinement and enhanced stability to deleterious MHD compared to conventional H-mode, and making hybrid scenarios attractive candidates for steady-state, reactor-relevant operation (see \cite{Turco_2023} and references therein).

This effort leverages DIII-D’s access to high-performance, well-diagnosed plasmas with a carbon wall and WEST’s capability for long-pulse operation with an actively cooled tungsten wall. In this context, DIII-D enables controlled, fully diagnosed studies of tungsten-induced core transport and rotation physics in preparation for future metal-wall operation, while WEST benefits from detailed physics interpretation of tungsten cooling effects that cannot be directly accessed owing to its more limited diagnostic capability, particularly for plasma rotation. The overall objectives and design of the coordinated experiments are presented in detail in Litaudon \textit{et al.}~\cite{Litaudon_2026}.  

Within this broader framework, the present work provides the first detailed transport studies of the effects of W cooling in DIII-D hybrid-like plasmas, operated with WEST-similarity constraints such as a matched radiated power fraction, shape and normalized core parameters. We examine two plasma conditions within the same hybrid scenario: a phase exhibiting strong core W radiation following controlled Laser Blow-Off (LBO) injection \cite{biwole_lbo_2026}, and a phase in which intrinsic carbon radiation dominates. In doing so, we isolate how enhanced cooling by W modifies ion and electron energy transport, momentum transport, and the resulting evolution of impurity profiles. Because hybrid scenarios on DIII-D exhibit MHD activity \cite{Turco_2024a,Turco_2024b}, including a benign $m/n=1/1$ sawtooth mode, we also assess the role of this activity in mediating impurity redistribution and temperature evolution during W cooling. 

The remainder of this paper is organized as follows. Section ~\ref{sec2_setup} describes the experimental setup, plasma parameters, and diagnostic systems. Section ~\ref{sec3_meas} presents the main experimental results, including changes in the background plasma profiles and their gyrokinetic predictions, as well as a power balance analysis comparing the different radiative phases. Section ~\ref{sec4_transport} provides detailed transport studies, covering heat, momentum, and impurity transport across the two radiative regimes, followed by an analysis of MHD activities. Section ~\ref{sec5_discuss} discusses the broader implications of the results, including consequences for L-H and H--L transitions, fast-ion confinement, and the relative roles of neoclassical and turbulent impurity transport, with relevance for interpreting WEST experiments. Section ~\ref{sec6_conclusion} summarizes the key results of this work. 
\section{Experimental setup}
\label{sec2_setup}

\begin{figure}[t]
\centering
\includegraphics[width=\columnwidth]{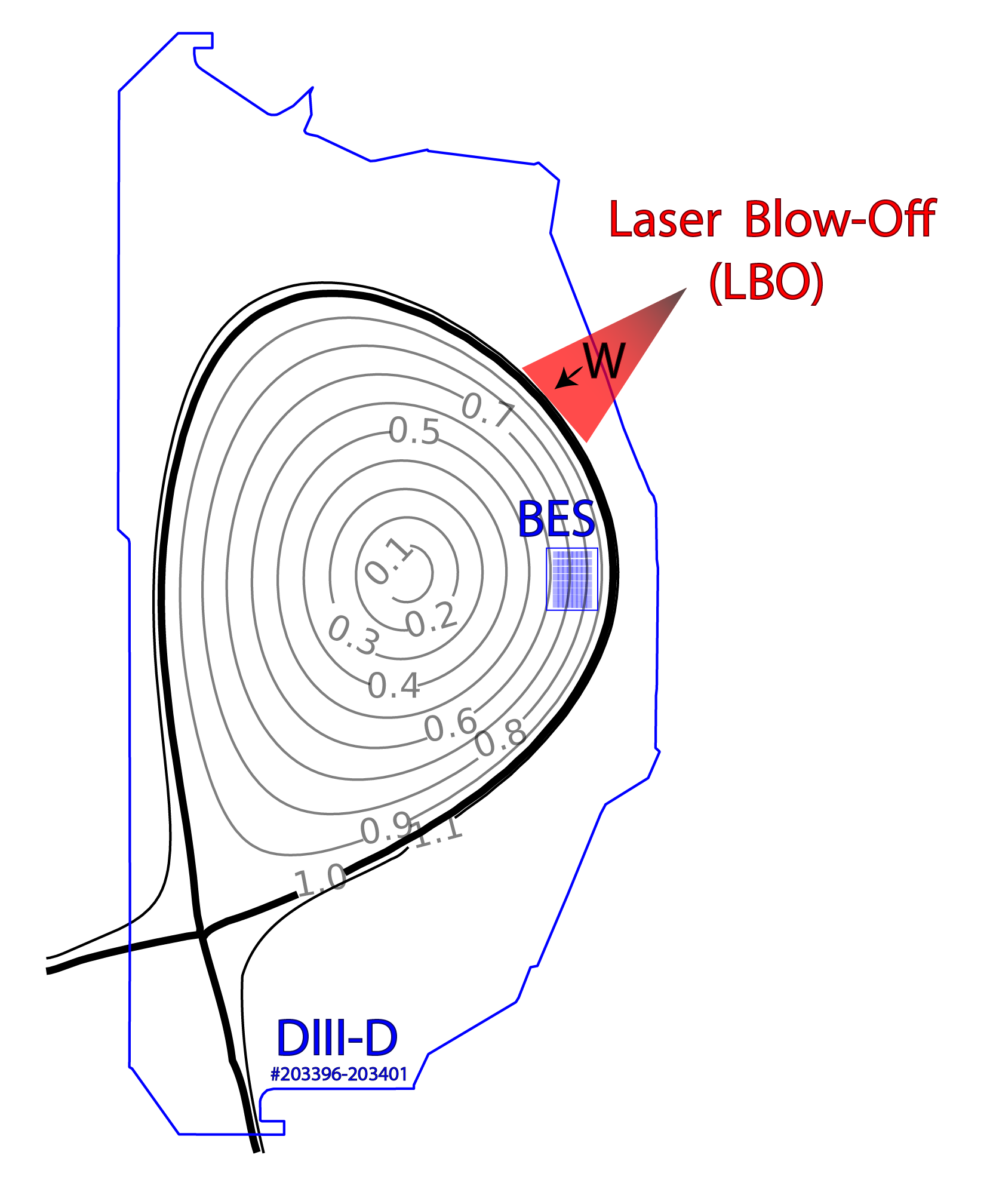}
\caption{ Poloidal cross-section of the plasma discharges of interest, showing the magnetic geometry matched to WEST parameters and the Beam Emission Spectroscopy (BES) channel locations spanning $0.6<\rho<0.9$ at the outboard midplane.  The geometry shown is representative of DIII-D discharges \#203396-\#203401. Tungsten is injected from the low-field side using the Laser Blow-Off (LBO) system, while BES provides localized turbulence measurements used in this study.}
\label{fig:fig1-cross_section}
\end{figure}

The experiment was carried out on the DIII-D tokamak. The discharges analyzed in this work are in the ITER ``hybrid-like'' regime \cite{Gormezano_2007,Petty_2015}, achieved under WEST-similarity constraints, including matched radiated power fraction ($f_{\mathrm{rad}} \lesssim 0.6$), dominant electron heating, matched magnetic geometry (weak-shear $q$-profile with $q_0>1$ and $q_{95}\approx5$, elongation $\kappa\approx1.3$, triangularity $\delta\approx0.45$, and X-point configuration), and comparable core-normalized parameters including $\beta$, electron Larmor radius $\rho_e^*\sim2.5\times10^{-4}$, collisionality $\nu_e^*\sim10^{-2}$, and $T_e/T_i\sim2$--$3$, and a core electron density of the order $n_{e0}\sim5\times10^{19}\,\mathrm{m}^{-3}$. No active density regulation was applied on DIII-D beyond a low-density safety threshold.

Discharge \#203401 is taken as the baseline case for the present analysis and is representative of the similarity set \#203396--\#203401, which share nearly identical plasma shape, magnetic configuration, and diagnostic coverage. High-quality profile measurements are obtained in \#203401, while BES turbulence measurements are available in \#203396. The corresponding poloidal cross-section for this set of discharges is shown in Fig. ~\ref{fig:fig1-cross_section}.

A more detailed description of the first similarity experiments between DIII-D and WEST can be found in \cite{Litaudon_2026}. This section defines the baseline discharge. The evolution of the core-normalized parameters as the radiated power fraction is matched on DIII-D constitutes a central result of this work and is discussed throughout the paper. While the WEST similarity discharges reported in \cite{Litaudon_2026} are operated in L-mode, the baseline DIII-D discharge analyzed here is in H-mode, under otherwise comparable similarity constraints.

The discharge was operated at constant plasma current $I_p \sim 0.7\,\mathrm{MA}$ and toroidal magnetic field strength $B_t \sim 2\,\mathrm{T}$. The plasma was heated with approximately $2\,\mathrm{MW}$ of electron cyclotron resonance heating (ECRH) and $3\,\mathrm{MW}$ of neutral beam injection (NBI), with the NBI torque nearly balanced globally, minimizing the net external momentum input $\sim 0 \pm 0.5\,\mathrm{N\,m}$. With such a low external momentum input, the formation of the rotation profiles is expected to rely primarily on intrinsic torque generation and corresponding transport, making these profiles particularly sensitive to changes in transport. NBI sources operated at energies below $81\,\mathrm{keV}$, providing mixed ion and electron heating depending on $T_e$, while ECRH was injected in X-mode at $110\,\mathrm{GHz}$ for absorption near the magnetic axis ($\rho_{\mathrm{ECH}} < 0.1$).

Controlled tungsten injection was achieved using the multi-pulse Laser Blow-Off (LBO) system \cite{biwole_lbo_2026}, in which a high-energy laser pulse ablates a thin tungsten coating, producing a directed plume of neutral W atoms entering the plasma from the low-field side (Fig. ~\ref{fig:fig1-cross_section}). Each LBO pulse injects approximately $20$--$30\,\mu\mathrm{g}$ of tungsten ($\sim 4\times10^{17}$ atoms), corresponding to a core concentration of the order $n_{\mathrm{W}}/n_e \sim 3$--$4\times10^{-4}$. The LBO system was operated at its maximum repetition rate of $10\,\mathrm{Hz}$, providing frequent injections that approximate a quasi-steady tungsten influx during the injection phase. The injected tungsten produced a clear transition from a low-radiation reference phase to a W-cooling phase with enhanced core radiation.

A comprehensive suite of diagnostics was used to characterize temperature, density, rotation, radiation, and impurity content. Electron temperature and density profiles were obtained from ECE and Thomson scattering, with CO$_2$ interferometry providing line-integrated density constraints.

Ion temperature and toroidal rotation were measured using charge-exchange recombination spectroscopy (CER) on C$^{6+}$. Multi-ion charge-exchange recombination (MICER) measurements were used to provide complementary information on main-ion temperature and density trends. Total tungsten density was inferred from 32-channel soft X-ray (SXR) arrays equipped with a $125\,\mu\mathrm{m}$ beryllium (Be) filter, with a temporal resolution of $\sim 1\,\mathrm{ms}$.

Profiles of radiated power densities were inferred from multi-chord bolometry measurements and tomographic inversion \cite{Odstrcil_2012,Odstr_il_2016}. The radiated power fraction $f_{\mathrm{rad}}$ is here defined as the ratio of the total radiated power inside the separatrix to the total absorbed heating power, including ohmic and auxiliary heating as well as fast-ion contributions. The effective charge $Z_{\mathrm{eff}}$ was inferred from visible bremsstrahlung measurements and is reported as a core-averaged quantity. Broadband density fluctuations were measured using beam emission spectroscopy (BES), with chord-viewing locations spanning $0.6 < \rho < 0.9$ (see Fig. ~\ref{fig:fig1-cross_section}).

\begin{figure}[t]
\centering
\includegraphics[width=\columnwidth]{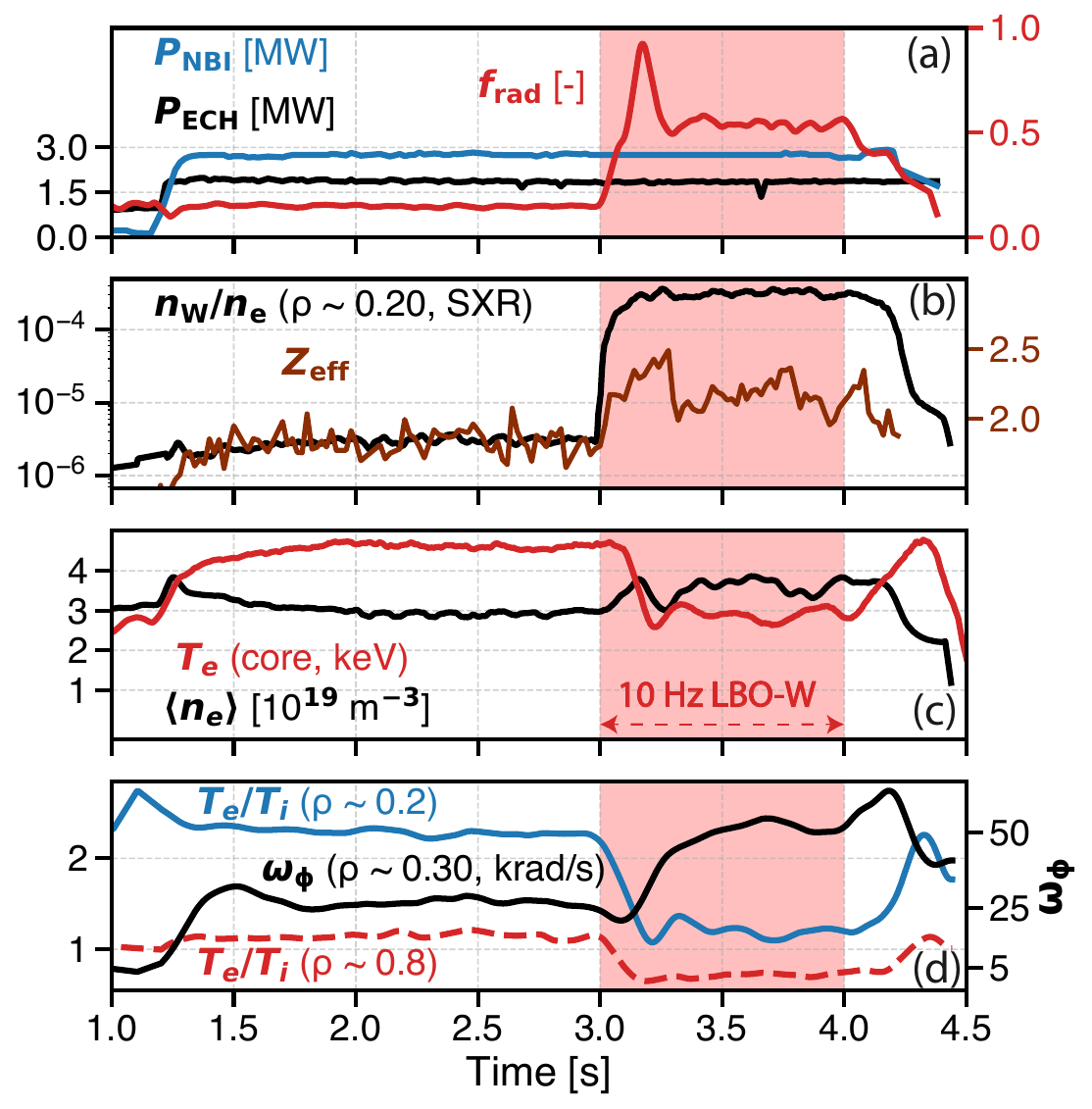}
\caption{Time traces from the baseline DIII-D discharge \#203401 showing the evolution of key plasma parameters. Panel (a) shows the applied NBI and ECH heating powers together with the core radiated power fraction. Panel (b) shows the inferred tungsten density from soft X-ray (SXR) measurements and the core-averaged effective charge $Z_{\mathrm{eff}}$. Panel (c) shows the core electron temperature (ECE) and the line-averaged electron density (Thomson scattering), while panel (d) shows the ion--electron temperature ratio and the toroidal rotation measured by CER at representative radii. The red-shaded interval marks the W-cooling phase following the controlled tungsten injection, while the preceding interval corresponds to the low-radiation reference phase dominated by intrinsic carbon radiation.}
\label{fig:fig2-time_trace}
\end{figure}

The time evolution of the main plasma parameters for this discharge is summarized in Fig.~\ref{fig:fig2-time_trace}. Panel (a) shows the applied NBI and ECH heating powers together with the core radiated power fraction, illustrating the transition into the W-cooling phase following the controlled LBO injection. Panel (b) shows the inferred tungsten density and core-averaged effective charge $Z_{\mathrm{eff}}$, indicating an increase in high-$Z$ impurity content during this interval. The evolution of the core electron temperature and line-averaged density is shown in panel (c), while panel (d) shows measurements of the ion--electron temperature ratio and toroidal rotation obtained from CER at representative radii. 

A notable feature of the W-cooling phase is the reduction of the electron-to-ion temperature ratio, which drops below unity at the outer plasma radius (see, for example, at $\rho \approx 0.8$). This reversal of the relative magnitude of $T_e$ and $T_i$ modifies the electron--ion heat exchange and has implications for the local power balance. In addition, the CER measurements show a significant increase in toroidal rotation during the W-cooling phase, consistent with the enhanced radiative losses and the associated changes in momentum transport. The red-shaded region marks the W-cooling phase, contrasted with the preceding reference phase dominated by intrinsic carbon radiation. These measurements define the two radiative conditions examined in the transport analysis that follows.
\section{Results}
In this section, we first present the experimentally measured changes in kinetic profiles induced by the LBO-W injection, then examine the corresponding power balance, and finally assess whether these profile changes can be reproduced by integrated gyrokinetic transport modeling using TGLF \cite{Staebler_2005} for turbulence and NEO \cite{Belli_2008} for neoclassical transport.
\label{sec3_meas}
\subsection{Observation of changes in background profiles}

\begin{figure*}[t]
\centering
\includegraphics[width=0.85\textwidth]{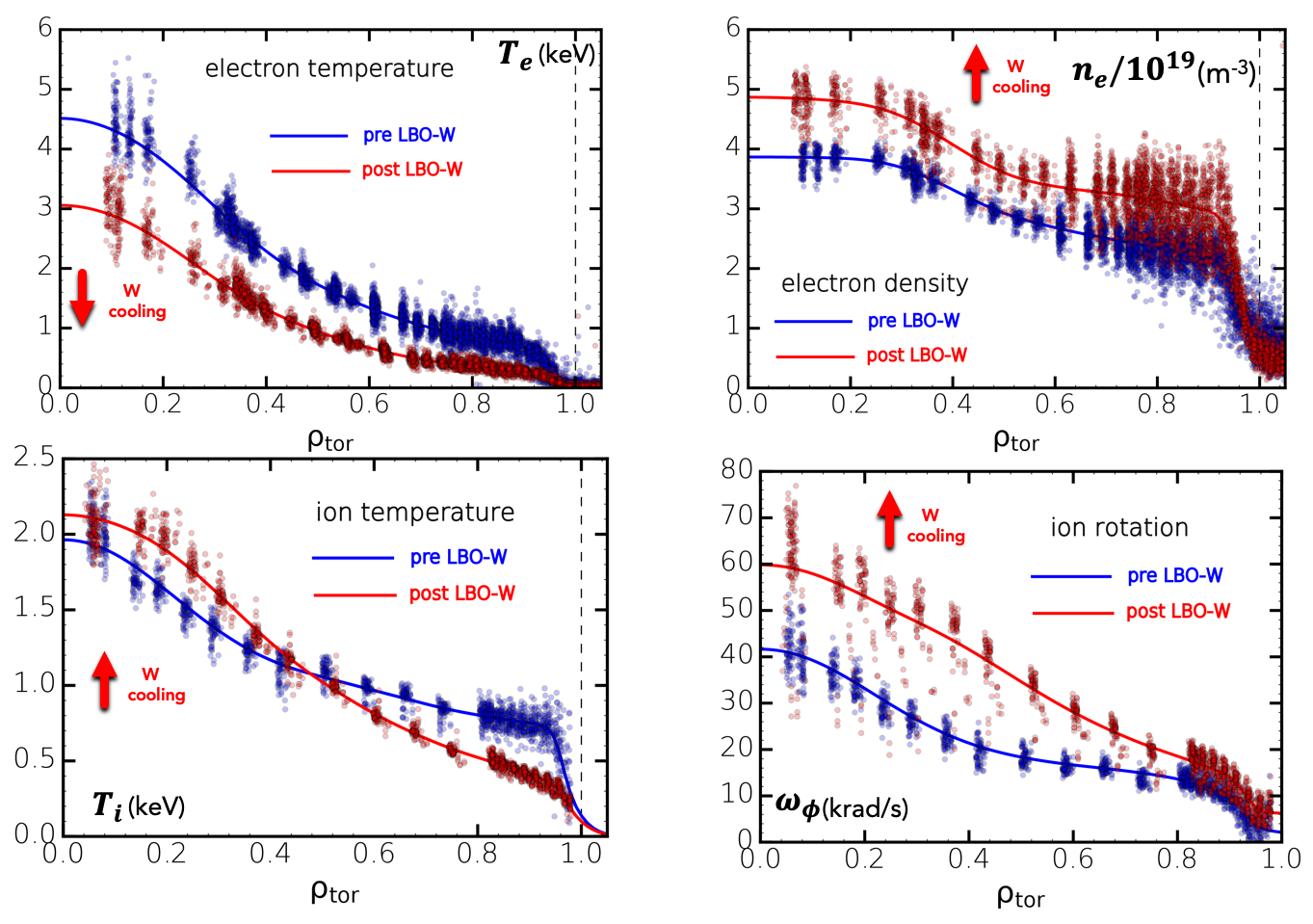}
\caption{ Measured kinetic profiles of the hybrid discharge before and after tungsten injection. Shown are the electron temperature $T_e$, electron density $n_e$, ion temperature $T_i$, and toroidal rotation $\omega_\phi$ as a function of the toroidal flux radius $\rho_{\mathrm{tor}}$. Blue points correspond to the reference phase (averaged over a $1\,\mathrm{s}$ interval before LBO), and red points correspond to the W-cooling phase (averaged over a $1\,\mathrm{s}$ interval after LBO). Solid curves represent fits to the averaged data. Tungsten injection leads to a strong reduction of $T_e$, an increase in $n_e$, a core peaking of $T_i$, and a significant increase in $\omega_\phi$.}
\label{fig:fig3-obs}
\end{figure*}

We now examine the measured plasma profiles before and after the LBO-W injection. The reference phase corresponds to a $\sim1\,\mathrm{s}$ stationary time window prior to the LBO injection (shown in blue in Fig.~\ref{fig:fig3-obs}), while the W-cooling phase corresponds to a comparable window during the enhanced-radiation period following tungsten injection (shown in red). Fig.~\ref{fig:fig3-obs} displays averaged profiles for these two phases. 

The electron temperature $T_e$ exhibits a pronounced decrease across the entire radius during the W-cooling phase, consistent with strong impurity-driven radiative cooling, with the core value $T_{e0}$ decreasing from $\sim4.5$ to $3\,\mathrm{keV}$. In contrast, the electron density $n_e$ increases over the whole profile, leading to a clear rise in the pedestal value. The ion temperature $T_i$ develops a modest core peaking during W cooling, with its central value increasing up to roughly $10\%$ (from $\sim2$ to $\sim2.2\,\mathrm{keV}$). The pedestal ion temperature decreases, though it remains higher than the electron pedestal temperature, a feature that has important consequences for the ion--electron energy exchange in the plasma edge. 

The toroidal rotation measured by CER also increases across the profile, with the most pronounced change near mid-radius, where the rotation doubles from $\omega_{\mathrm{tor}}\sim20$ to $\omega_{\mathrm{tor}}>40\,\mathrm{krad/s}$ at $\rho\approx0.4$. In terms of deuterium ion Mach number, the values remain modest, with $M_D\approx0.16$ (core-averaged, $\rho<0.3$) and $M_D\approx0.11$ at $\rho\sim0.8$ during the pre-tungsten phase, increasing to $M_D\approx0.23$ (core) and $M_D\approx0.18$ at $\rho\sim0.8$ during the W-cooling phase. It is worth noting that this increase, while substantial and leading to high rotation values for a discharge with very low NBI torque and central ECRH, yields absolute rotation levels that remain within the range generally observed on DIII-D, where values of $\sim60\,\mathrm{krad/s}$ ($\sim100\,\mathrm{km/s}$) are attainable.

\subsection{Power balance analysis}
\begin{figure*}[t]
\centering
\includegraphics[width=0.85\textwidth]{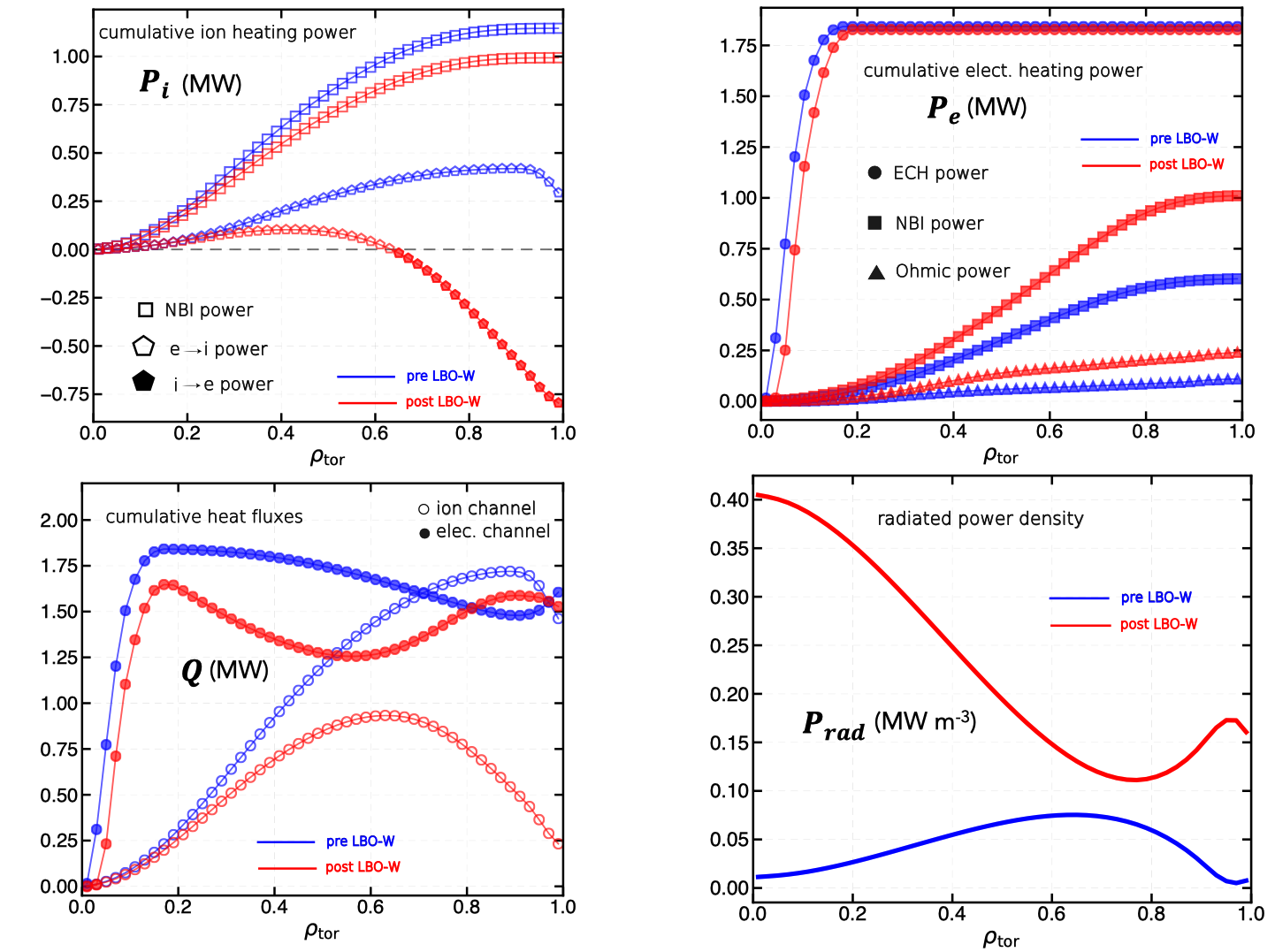}
\caption{ Power profiles for the reference (pre-LBO-W, blue) and W-cooling (post-LBO-W, red) phases of the discharge. All profiles are time-averaged over the intervals $2.3$--$2.9\,\mathrm{s}$ (pre-LBO-W) and $3.3$--$3.9\,\mathrm{s}$ (post-LBO-W). Cumulative ion heating, electron heating, and conductive/convective heat fluxes are shown as volume integrals from the magnetic axis to the normalized toroidal flux radius $\rho_{\mathrm{tor}}$. The radiated power density $P_{\mathrm{rad}}$ is shown as a local volumetric loss term derived from bolometer measurements using tomographic inversion, while the heating and heat flux terms are obtained from TRANSP power balance calculations.}
\label{fig:fig4-heats}
\end{figure*}

A detailed power balance was computed for both the reference and W-cooling phases using the TRANSP code \cite{Pankin_2025,Grierson_2018}. The corresponding power-flow diagrams are shown in Fig.~\ref{fig:fig4-heats}.

In the reference phase (time-averaged over $2.3$--$2.9\,\mathrm{s}$), the total injected NBI power of $3.3\,\mathrm{MW}$ is partitioned primarily to the ions, with $P_{\mathrm{NBI},i}\approx1.15\,\mathrm{MW}$ and $P_{\mathrm{NBI},e}\approx0.6\,\mathrm{MW}$. The remaining injected beam power is primarily lost through fast-ion charge exchange with background neutrals, with smaller contributions from beam shine-through and prompt orbit losses, while a fraction is transferred to the bulk plasma through fast-ion thermalization. ECH supplies $1.8\,\mathrm{MW}$ directly to the electrons, while the ohmic contribution is small ($P_{\mathrm{OH}}\approx0.1\,\mathrm{MW}$). The collisional energy exchange from electrons to ions is $P_{ei}\approx0.25\,\mathrm{MW}$. The resulting radiated power is $P_{\mathrm{rad}}\approx0.6\,\mathrm{MW}$, obtained by volume-integrating the radiated power density profile shown in Fig.~\ref{fig:fig4-heats}. The outward conductive/convective heat fluxes are $Q_i\approx1.45\,\mathrm{MW}$ and $Q_e\approx1.6\,\mathrm{MW}$, consistent with a well-confined, low-radiation H-mode DIII-D hybrid plasma. 

We note that the radiated power density used as input for the transport modeling resulted from tomographic reconstruction of the bolometric data, following the methods described in Refs.~\cite{Odstrcil_2012,Odstr_il_2016}. We also note that the pedestal values of the collisional electron--ion energy exchange term carry uncertainty, since they depend on the measured ion temperature $T_i$, which is less well constrained in this region of higher collisionality.

During the W-cooling phase (time-averaged over $3.3$--$3.9\,\mathrm{s}$), several major changes occur in the power balance. The NBI power redistributes nearly evenly between channels, with both the ion and electron heating converging to $P_{\mathrm{NBI},i}\approx P_{\mathrm{NBI},e}\approx1\,\mathrm{MW}$. This shift in partitioning is expected and is governed by the change in the critical velocity $v_{\mathrm{crit}}$: as tungsten radiation cools the electrons, the reduction in $T_e$ increases the ratio $v_{\mathrm{beam}}/v_{\mathrm{crit}}\propto T_e^{-1/2}$, causing a larger fraction of the beam power to be deposited directly to the electrons rather than the ions. This mechanism controls the relative power deposition between species during the cooling phase.

The ohmic heating increases by a factor of $\sim2.5$, reaching $P_{\mathrm{OH}}\approx0.25\,\mathrm{MW}$, reflecting the higher plasma resistivity at lower $T_e$ and higher $Z_{\mathrm{eff}}$. The collisional energy exchange reverses direction, with $P_{ei}\approx0.8\,\mathrm{MW}$ now flowing from ions to electrons, a consequence of the reduction of $T_e/T_i$ to below unity in the outer plasma radius. Most significant, but expected, is the rise in volume-integrated radiated power to $P_{\mathrm{rad}}\approx2.2\,\mathrm{MW}$, more than a factor of three above the reference phase. Crucially, the ion heat flux collapses from $Q_i\approx1.4$ to $\approx0.25\,\mathrm{MW}$, while the electron heat flux decreases more moderately to $Q_e\approx1.45\,\mathrm{MW}$.

This asymmetric response, i.e., sharp suppression of ion heat transport with only a modest reduction in electron heat transport, is consistent with a significant change in the underlying turbulence regime and in the electron--ion collisional coupling during the W-cooling phase. These effects will be examined in detail in Sec.~\ref{sec4_transport}, and may help explain both the rise in toroidal rotation and the modest peaking of the ion temperature, whereas the present power-balance analysis already captures the primary changes in the electron profiles. In the next subsection, we use gyrokinetic-based transport modeling to test whether background profile reproduction further constrains the turbulence changes responsible for the observed evolution.

\begin{figure}[t]
\centering
\includegraphics[width=\columnwidth]{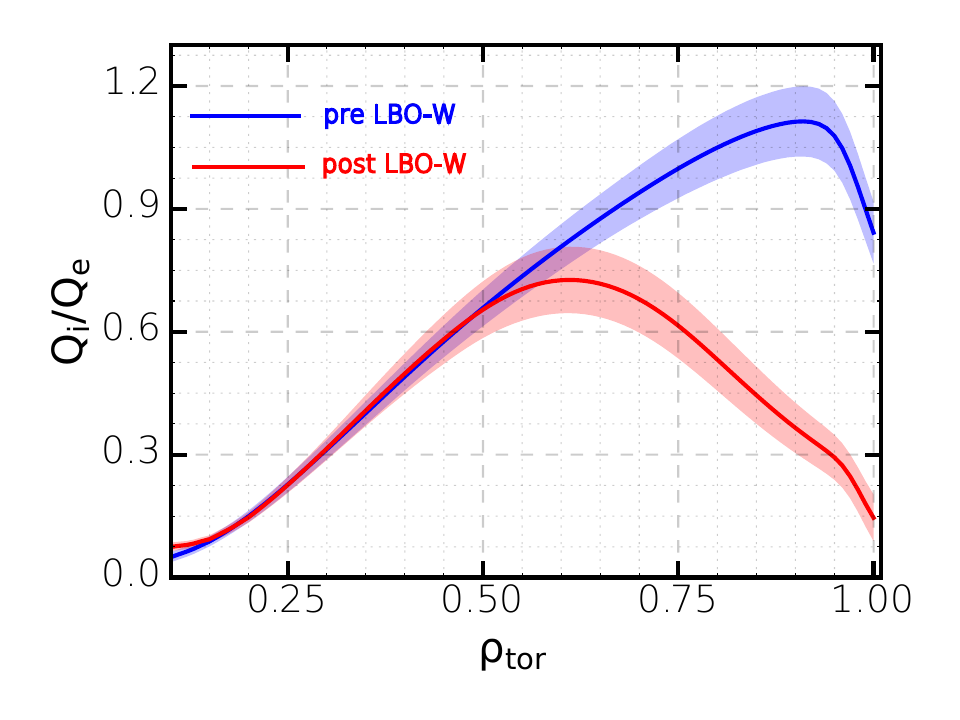}
\caption{Ratio of ion to electron heat flux $Q_i/Q_e$ before (blue) and after (red) the LBO-W injection, obtained from TRANSP power-balance analysis and time-averaged over the same intervals as Fig.~\ref{fig:fig4-heats}. Following the W injection, $Q_i/Q_e$ decreases across the outer plasma region, indicating a shift toward more electron-dominated heat transport. This behavior reflects the strong suppression of ion heat transport during the W-cooling phase while the electron heat flux decreases more modestly.}
\label{fig:fig5-ratio}
\end{figure}

The redistribution of heat transport is further illustrated by the ratio $Q_i/Q_e$ shown in Fig.~\ref{fig:fig5-ratio}. Following the W injection, $Q_i/Q_e$ decreases sharply across the outer radius, indicating a shift toward more electron-dominated heat transport. This behavior is consistent with the power-balance analysis, which shows a strong suppression of ion heat flux while the electron channel decreases more modestly during the W-cooling phase. The reduction of $Q_i/Q_e$ also reflects the regime where $T_e<T_i$ in the outer plasma region, allowing ions to temporarily act as an energy reservoir for electrons during radiative cooling.

\subsection{Gyrokinetic-based prediction of background profile changes}

To determine whether the observed evolution of the kinetic profiles during the W-cooling phase can be explained by changes in turbulent and neoclassical transport, we performed integrated modeling using TGYRO \cite{Candy_2009}, with turbulent fluxes computed by TGLF \cite{Staebler_2005} and neoclassical fluxes computed by NEO \cite{Belli_2008}. The simulations evolve $n_e$, $T_e$, and $T_i$ to match the experimental power balance, using radiated-power density profiles from bolometer tomography. Two electromagnetic saturation rules (SAT0 and SAT2) are considered, together with a sensitivity case in which tungsten radiation is removed.

\begin{figure*}[t]
\centering
\includegraphics[width=0.95\textwidth]{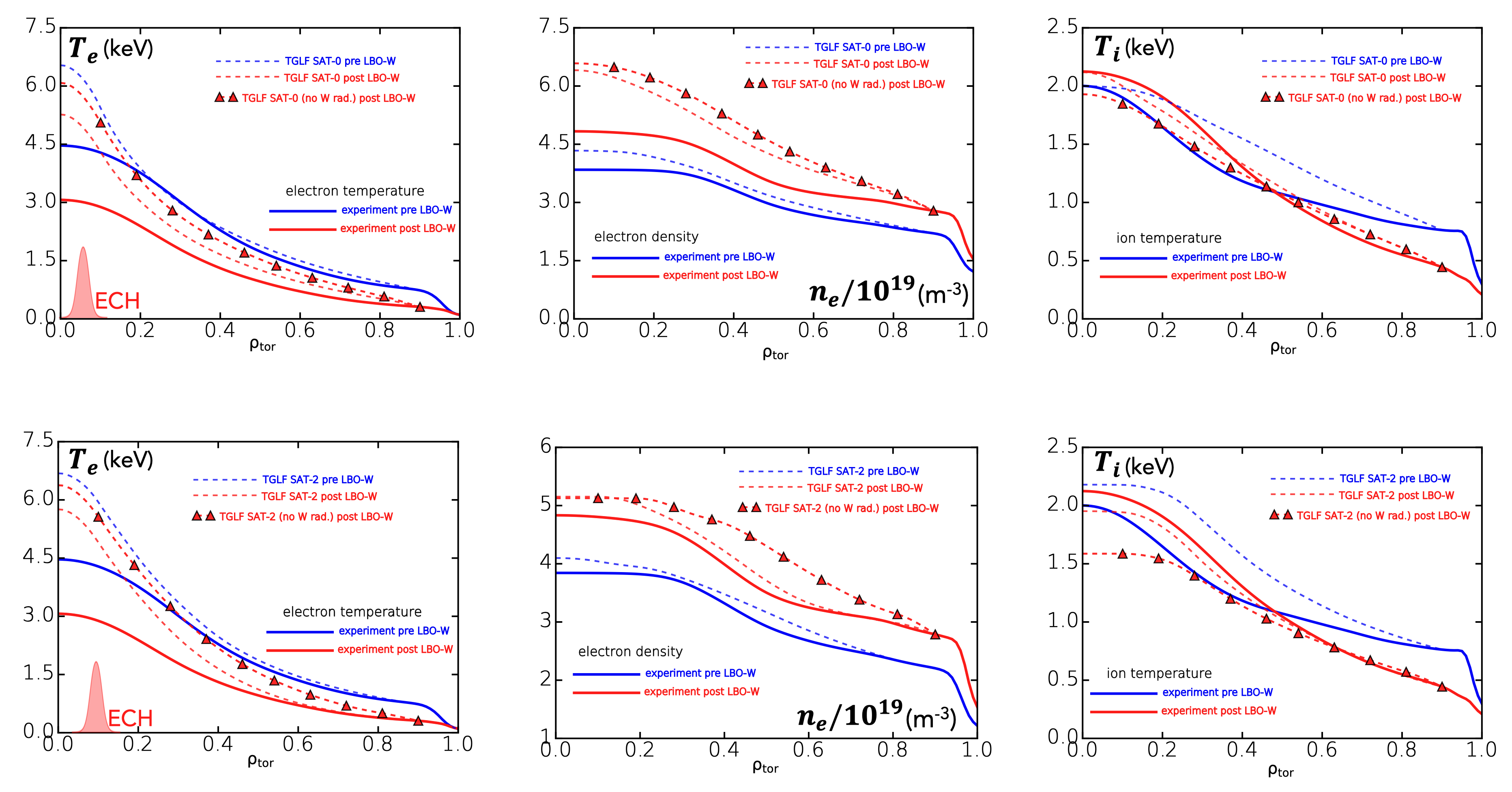}
\caption{Comparison of experimental kinetic profiles and TGYRO predictions for the pre-LBO and W-cooling phases using electromagnetic TGLF saturation rules (SAT0 and SAT2). Shown are electron temperature, electron density, and ion temperature versus normalized toroidal flux radius. Solid lines denote experimental profiles, dashed lines indicate TGLF predictions including tungsten radiation, and triangle markers correspond to simulations excluding tungsten radiation. Both models reproduce the global reduction of $T_e$ during W cooling, while overestimating the core temperature. The modest $T_i$ peaking is captured by SAT0 but not SAT2. Electron density is generally overpredicted in the W-cooling phase.}
\label{fig:fig6-tgyro}
\end{figure*}

Fig.~\ref{fig:fig6-tgyro} compares the experimental profiles with TGYRO predictions for the reference and W-cooling phases. Both models reproduce the global reduction of electron temperature following tungsten injection, although they systematically overestimate the core temperature relative to measurements. Removing tungsten radiation degrades the agreement, with higher predicted post-LBO temperatures, indicating that impurity radiation contributes significantly to the observed cooling. However, the change remains moderate, showing that the temperature evolution is not driven solely by radiation but also by modifications of turbulent transport.

The ion-temperature response depends more strongly on the saturation rule. The modest core $T_i$ peaking observed experimentally during the W-cooling phase is reproduced by the SAT0 model, whose predicted $T_{i0}$ matches the measurement. In contrast, SAT2 does not reproduce the increase in core ion temperature. Removing tungsten radiation further worsens the $T_i$ agreement, indicating that impurity-induced cooling influences ion transport indirectly through changes in background gradients and power balance.

The electron density evolution is less well captured. Both models overestimate the density increase during the W-cooling phase, particularly SAT0, while SAT2 provides somewhat improved but still elevated predictions relative to experiment.

Electrostatic variants of the models yield similar qualitative behavior, indicating that electromagnetic effects do not qualitatively change the predicted profile evolution. Toroidal rotation was not evolved in the baseline simulations shown here. In additional cases where rotation evolution was enabled using the radial electric field inferred from CER measurements, flux-matched solutions including angular-momentum transport did not reproduce the observed core rotation, particularly in the W-cooling phase. Rotation physics is therefore addressed separately in Sec.~\ref{sec4_transport}.

The modest ion-temperature peaking observed during the W-cooling phase is reproduced by the SAT0 electromagnetic model without invoking impurity dilution or additional ion-driven modes. This suggests that the dominant mechanism responsible for the profile evolution is a modification of the existing turbulence, consistent with turbulence stabilization during radiative cooling. While dilution and other effects may contribute, they are not required to explain the observed $T_i$ peaking in this case.

Taken together, these results indicate that the observed profile evolution arises from the combined effects of impurity-driven radiative cooling and modified turbulent transport. While the reduced models reproduce the overall trends, differences between saturation rules and persistent discrepancies in density and rotation highlight limitations of the predictive capability in strongly radiating regimes.

\section{Transport studies}
In this section, we investigate the transport mechanisms underlying the observed profile evolution, focusing first on heat and momentum transport through linear stability analysis, quasilinear sensitivity scans, and turbulence measurements, before examining the implications for impurity transport.
\label{sec4_transport}
\subsection{Heat and momentum transport}
\subsubsection{Linear stability}
The linear gyrokinetic analysis presented here is based on the experimentally measured profiles described in Sec.~\ref{sec3_meas}. Linear stability calculations were performed using TGLF and the CGYRO gyrokinetic code \cite{Candy_2016}, which is used here in linear mode to compute growth rates and identify the dominant microinstabilities for the measured plasma conditions.

\begin{figure*}[t]
\centering
\includegraphics[width=0.85\textwidth]{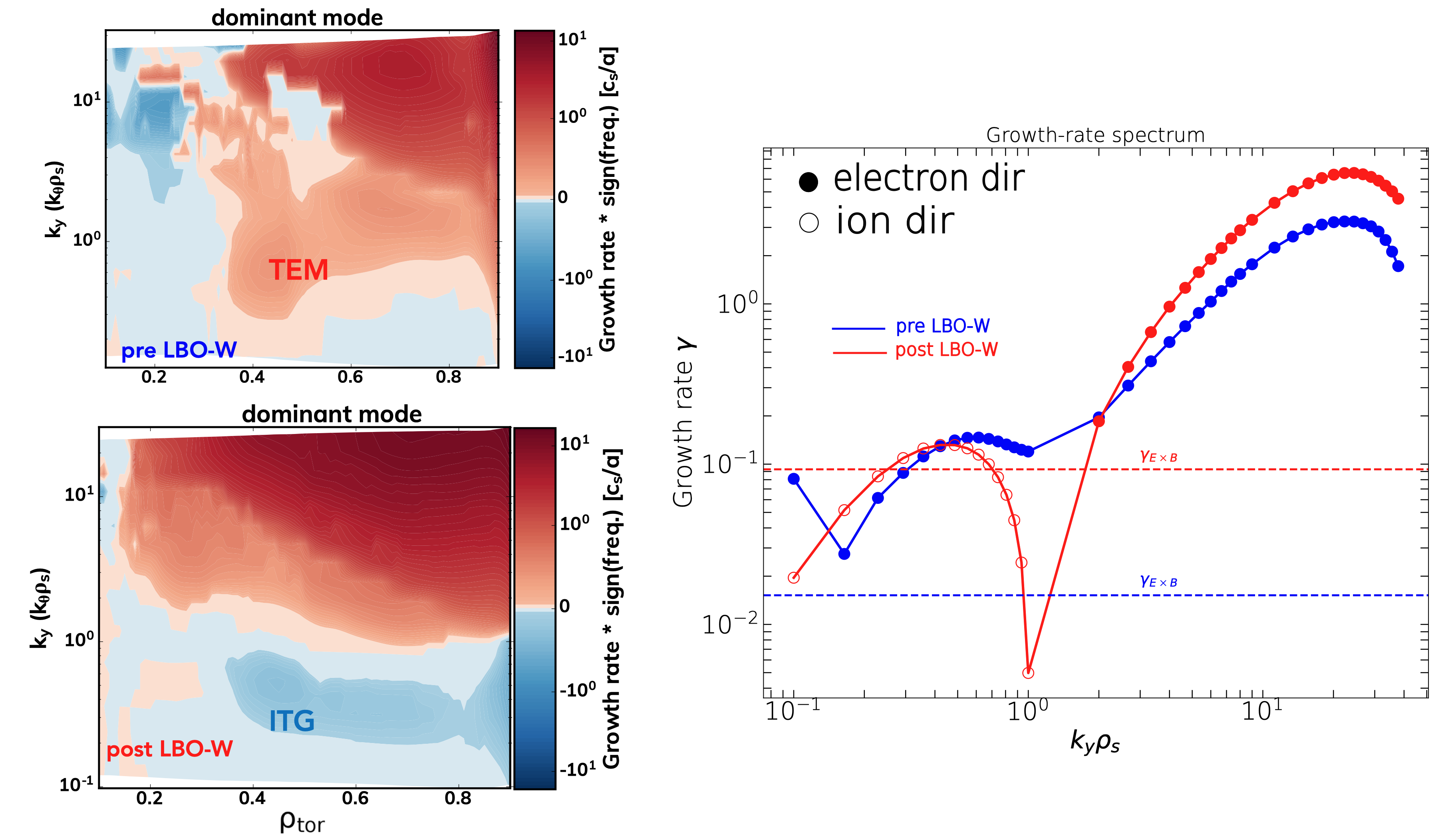}
\caption{TGLF (left) and linear CGYRO (right) stability analysis before and after the LBO-W injection. Left: Dominant linear microinstability as a function of radius and binormal wavenumber $k_y\rho_s$. Before LBO-W (top), the turbulence is predominantly TEM-driven across most of the radius, whereas after LBO-W (bottom) the dominant mode switches to ITG over a broader region. Colors indicate the signed growth rate, with red (blue) corresponding to electron (ion) direction propagation. 
Right: Growth-rate spectra at mid-radius $\rho=0.6$, comparing the pre- and post-LBO-W phases. Solid markers denote electron-direction modes, and open markers denote ion-direction modes. Horizontal dashed lines indicate the local $\ExB$ shear rate for each phase at the given radius. The LBO-W injection favors ITG modes, consistent with the mode-map transition in the left panels, although the ion-scale growth rate becomes closer to the $\ExB$ shear rate.}
\label{fig:fig7-linear}
\end{figure*}

Fig.~\ref{fig:fig7-linear} provides the starting point for understanding the transport changes induced by the W injection. Before LBO-W, the plasma state is characterized by mixed TEM/ITG turbulence across the radius, with growth rates well above the local $\mathbf{E}\times\mathbf{B}$ shearing rate. After the W-driven cooling, the linear spectrum shifts toward ITG-dominated modes with growth rates now approaching the shearing rate in the range $0.1 < k_y\rho_s < 1$. This places the plasma in a regime where the increase in $\mathbf{E}\times\mathbf{B}$ shear is able to suppress a significant fraction of the remaining ion-scale turbulence. The increase in $\mathbf{E}\times\mathbf{B}$ shearing rate illustrated in Fig.~\ref{fig:fig7-linear} (right panel) for radius $\rho = 0.6$ is valid for all plasma radii, as the toroidal rotation increases across the radius in the W-cooling phase.

\subsubsection{Turbulence sensitivity scan}
\begin{figure*}[t]
\centering
\includegraphics[width=0.85\textwidth]{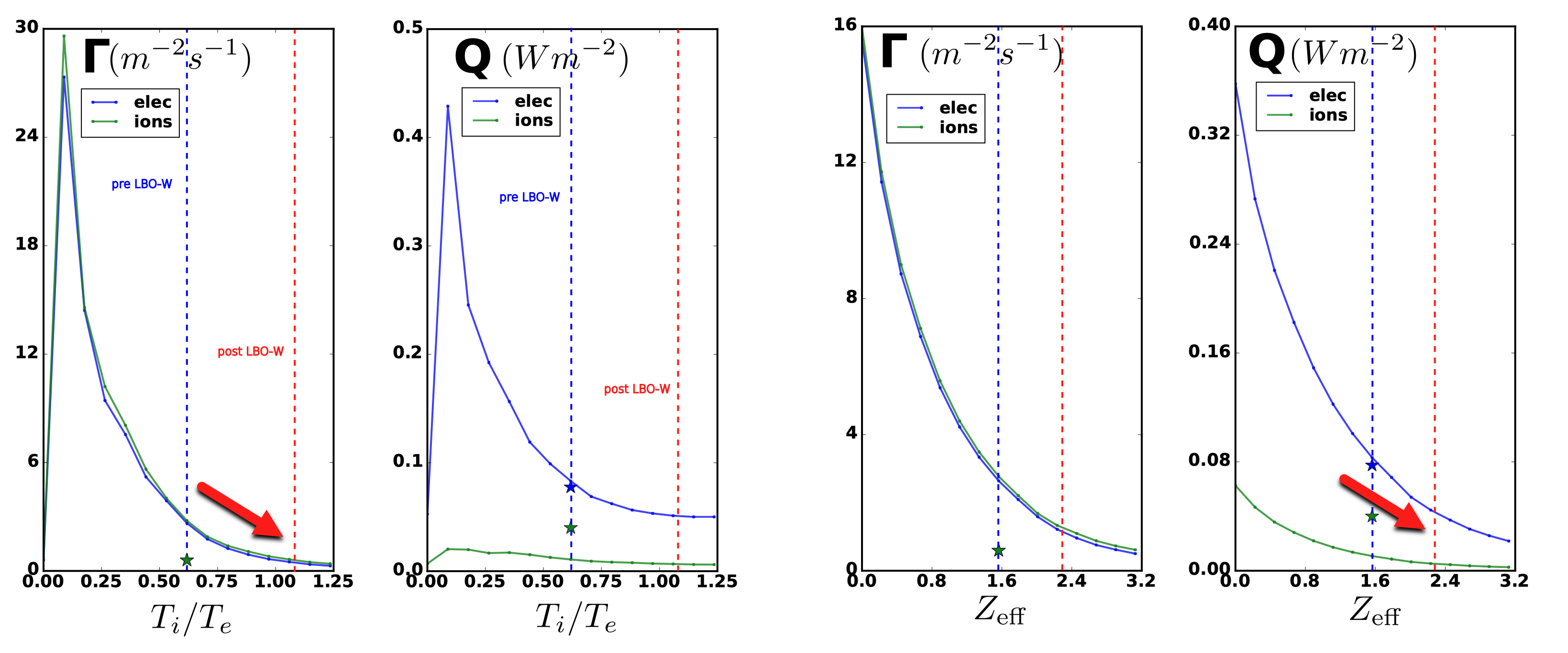}
\caption{Quasilinear TGLF sensitivity scans of the electron and ion heat fluxes $Q$ at $\rho = 0.5$. Left: sensitivity to the temperature ratio $T_i/T_e$. The blue and red dashed vertical lines indicate the experimental pre- and post-LBO-W values, respectively, showing that the W-cooling phase shifts the plasma toward larger $T_i/T_e$. Right: sensitivity to the effective charge $Z_{\mathrm{eff}}$. The blue and red dashed vertical lines indicate the corresponding experimental values before and after LBO-W, showing that the W-cooling phase also shifts the plasma toward higher $Z_{\mathrm{eff}}$. Star symbols mark the experimental heat flux magnitudes. In both scans, the post-LBO-W state moves toward a regime of reduced predicted heat flux, consistent with turbulence stabilization during tungsten radiative cooling.}
\label{fig:fig8-tglfscan}
\end{figure*}

The quasilinear TGLF and CGYRO sensitivity scans shown in Fig.~\ref{fig:fig8-tglfscan} were performed using the SAT0 electromagnetic saturation rule, which provided the best agreement with the observed ion-temperature peaking in Sec.~3.3. These scans of $T_e/T_i$ and $Z_{\mathrm{eff}}$ further clarify why turbulence stabilizes. The W injection cools the electrons, thereby decreasing $T_e/T_i$ and raising the effective charge. Both trends reduce the turbulent fluxes, stabilizing TEMs and driving the system deeper into an ITG-dominated state with lower overall transport. These scans show that the experimentally observed operating points move into regions of reduced particle, heat, and momentum fluxes, consistent with the transition from mixed TEM/ITG to a more ITG-stable regime.

We note that the stabilization of TEM turbulence with decreased $T_e/T_i$ is a well-understood phenomenon. On DIII-D, previous work has experimentally verified the opposite behavior, namely, an increase in turbulent activity when $T_e/T_i$ is increased through electron cyclotron resonance heating (ECRH) \cite{Thome_2021,Petty_1999,Ernst_2016,Pinsker_2015,McKee_2000}. 

The observations reported here complement these studies by demonstrating that electron cooling produced by high-$Z$ radiative losses decreases turbulence, effectively counteracting the usual ECH-induced destabilization.

Importantly, this stabilization is not driven by main-ion dilution. A natural question is whether the observed reduction of turbulent transport could instead arise from dilution caused by tungsten injection. This does not seem to be the case for high-$Z$ impurities such as tungsten. From quasi-neutrality, the main-ion fraction satisfies $n_D/n_e = (Z - Z_{\mathrm{eff}})/(Z - 1)$. 

For a representative tungsten charge state in the plasma core with $Z \simeq 54$, and $Z_{\mathrm{eff}} \simeq 2$--$3$ as in the present experiment, $Z \gg Z_{\mathrm{eff}}$, which yields $n_D/n_e \approx 0.96$, indicating negligible deuterium dilution even as tungsten concentration increases; the observed turbulence stabilization therefore cannot be attributed to dilution.

The quasilinear scans predict reductions in both electron and ion heat fluxes as $T_i/T_e$ increases and $Z_{\mathrm{eff}}$ rises. This behavior is qualitatively consistent with the power-balance analysis, which shows a strong reduction in ion heat flux and a more modest decrease in the electron channel during the W-cooling phase. The stronger suppression of ion transport suggests that turbulence stabilization primarily affects ion-scale transport.  The scans isolate the effect of changing $T_i/T_e$ and $Z_{\mathrm{eff}}$ at fixed $\ExB$ flow shear. In the experiment, however, the W-cooling phase is also accompanied by a significant increase in $\ExB$ shear, as shown by the linear stability analysis in Fig.~\ref{fig:fig7-linear}. This additional shear preferentially suppresses ion-scale turbulence, explaining why the power-balance analysis shows a much stronger reduction in ion heat flux than suggested by the quasilinear scans alone.

\subsubsection{Turbulent diffusivity}
\begin{figure}[t]
\centering
\includegraphics[width=\columnwidth]{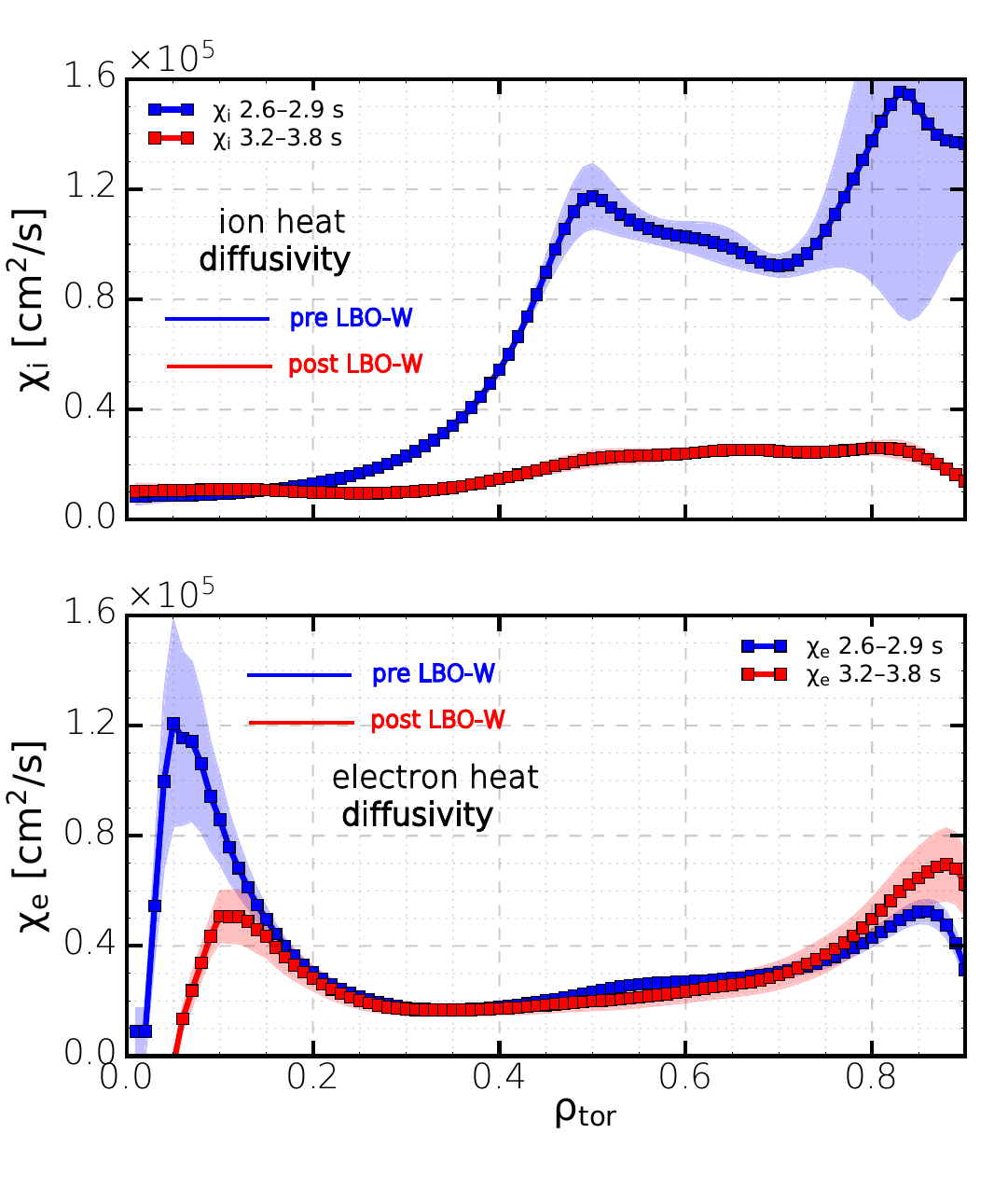}
\caption{Ion and electron heat diffusivities before and after the LBO-W injection.
Left panels: Radial profiles of the ion ($\chi_i$, top) and electron ($\chi_e$, bottom) heat diffusivities obtained from power-balance analysis. Blue curves correspond to the pre--LBO-W phase and red curves to the post--LBO-W high-radiation phase, with shaded regions indicating uncertainties, time-averaged over $2.6$--$2.9\,\mathrm{s}$ and $3.2$--$3.8\,\mathrm{s}$. $\chi_i$ decreases following LBO-W across all radii, while $\chi_e$ decreases mainly around the ECH deposition location.
Right panels: Time evolution of $\chi_i$ and $\chi_e$ at selected radial locations ($\rho = 0.1, 0.3, 0.5, 0.7, 0.9$). The figure highlights the increase in electron diffusivity $\chi_e$ at radius $\rho=0.9$ during the high W-cooling phase.}
\label{fig:fig9-diff}
\end{figure}

The inferred ion and electron heat diffusivities (Fig.~\ref{fig:fig9-diff}), obtained from TRANSP power-balance analysis, confirm this stabilization. The ion diffusivity $\chi_i$ drops strongly toward the edge during the W-cooling phase, reflecting the quenching of residual ITG activity as the $\ExB$ shear becomes more effective. The electron diffusivity $\chi_e$ decreases mainly around the ECH deposition radius, with little change in the core and a mild rise near the very edge. The change in the core $\chi_e$ is readily explained by the effects of W cooling in the core (effectively counteracting ECH), while the rise in edge diffusivity necessitates heat-flux analysis, which follows.

\subsubsection{Experimental validation of turbulence reduction}

\begin{figure}[t]
\centering
\includegraphics[width=0.8\columnwidth]{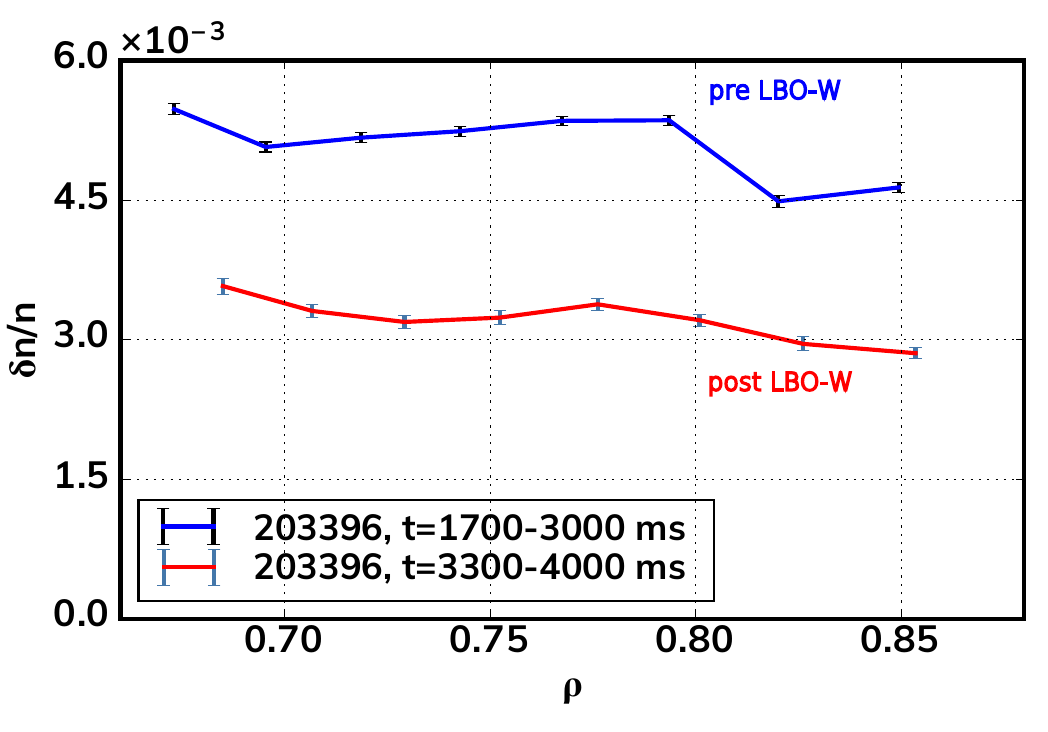}
\caption{Radial profile of normalized BES density fluctuation amplitude $\delta n/n$ before and after the W injection. Fluctuations are integrated over the frequency band $f = 20$--$120\,\mathrm{kHz}$. The pre-LBO phase ($1700$--$3000\,\mathrm{ms}$, blue) shows higher fluctuation levels across the measurement region $\rho \approx 0.7$--$0.85$, while the stationary post-LBO phase ($3300$--$4000\,\mathrm{ms}$, red) exhibits a clear reduction of density fluctuations. The conversion from intensity fluctuations $\delta I/I$ to density fluctuations uses Thomson scattering measurements of $n_e$ and $T_e$. The BES signal is sensitive to total plasma density fluctuations and does not distinguish between electron and ion contributions.}
\label{fig:fig10-dn}
\end{figure}

\begin{figure}[t]
\centering
\includegraphics[width=\columnwidth]{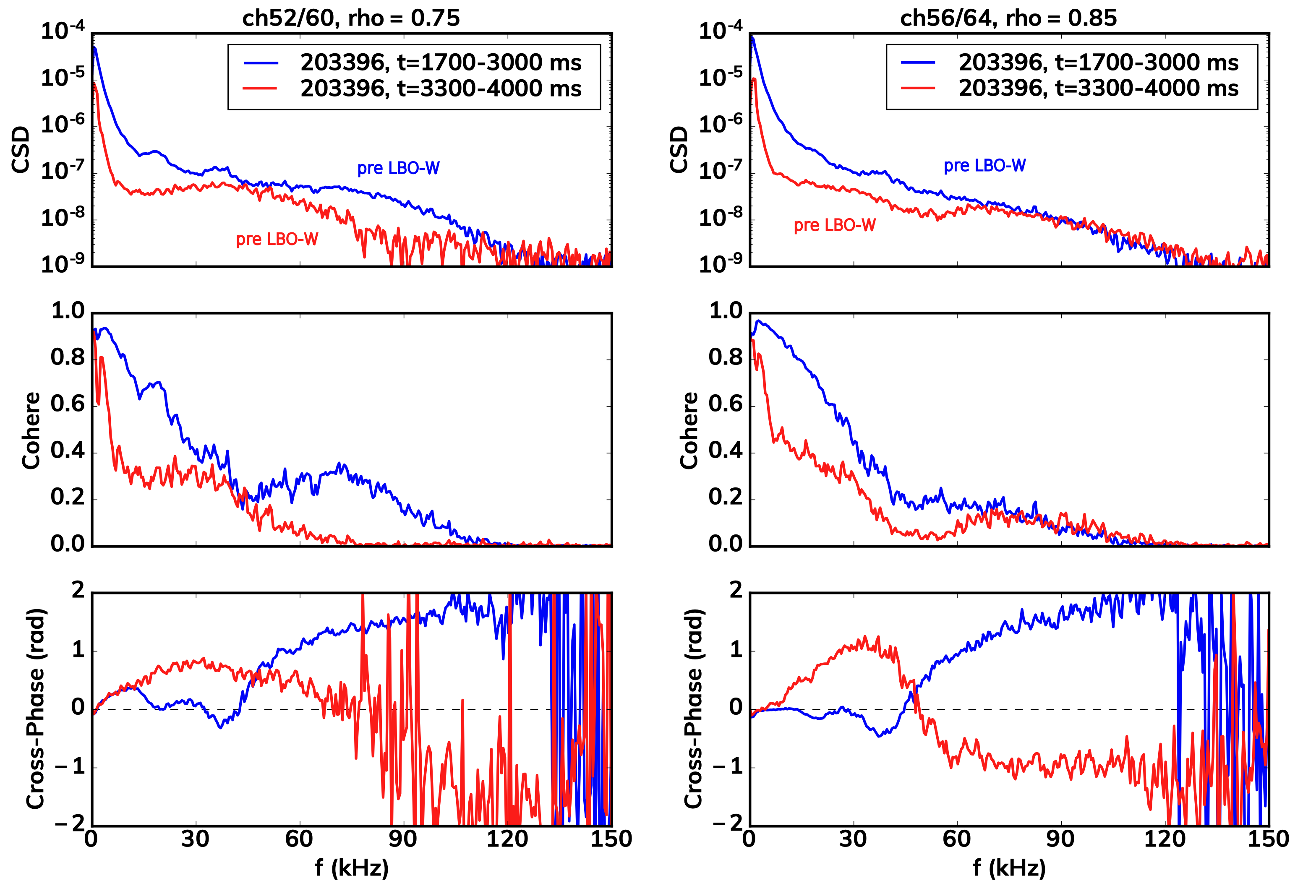}
\caption{Cross-spectral density (top), coherence (middle), and cross-phase (bottom) between poloidally separated BES channels at $\rho \approx 0.75$ (left) and $\rho \approx 0.85$ (right), before ($1700$--$3000\,\mathrm{ms}$, blue) and after ($3300$--$4000\,\mathrm{ms}$, red) the W injection. The analysis uses the frequency range $f = 20$--$120\,\mathrm{kHz}$. The post-LBO phase shows reduced fluctuation power and coherence across the ion-scale frequency range, indicating a weakening of broadband turbulence. The cross-phase evolution suggests a modification of the dominant fluctuation characteristics at the outer radii during the W-cooling phase.}
\label{fig:fig11-csd}
\end{figure}

Direct experimental evidence of reduced turbulence is provided by BES measurements (Figs.~\ref{fig:fig10-dn} and \ref{fig:fig11-csd}) obtained in a repeat discharge with similar tungsten radiation and plasma parameters. The fluctuation amplitude in the $20$--$120\,\mathrm{kHz}$ band decreases significantly after the W injection across the BES measurement region ($\rho \approx 0.7$--$0.85$), indicating reduced broadband density fluctuations.  

The reduction is observed during the stationary post-LBO phase ($3300$--$4000\,\mathrm{ms}$), avoiding the transient radiation spike immediately following injection. The BES signal is converted from intensity fluctuations to density fluctuations using Thomson-scattering measurements of $n_e$ and $T_e$. Because the BES diagnostic is sensitive to both electron and ion density perturbations, the measurements represent total plasma density fluctuations.  

Cross-spectral analysis at $\rho \approx 0.75$ and $\rho \approx 0.85$ shows a reduction in fluctuation power and coherence following the W injection, together with a modification of the cross-phase in the $20$--$120\,\mathrm{kHz}$ band. These observations indicate a weakening of broadband ion-scale turbulence in the outer region.  

These measurements are consistent with the gyrokinetic interpretation of turbulence stabilization during the W-cooling phase. While linear CGYRO does not predict fluctuation amplitudes, the predicted reduction of ion-scale growth rates and the increased $\ExB$ shear are qualitatively consistent with the observed decrease in BES fluctuation levels.

\subsubsection{Momentum transport}
These modifications of the turbulence directly impact the formation of the rotation profiles. Due to the overall low input torque, intrinsic momentum sources and transport are dominant in setting the rotation self-consistently. For a more detailed description, see the review by Angioni \textit{et al.}~\cite{Angioni_2012}.

Building on recent progress in validating momentum transport predictions, a reduced-rotation model, an upgraded version of that of Zimmermann \textit{et al.}~\cite{Zimmermann_2024}, is applied to the experimental scenario discussed herein. The reduced rotation model uses the experimentally measured kinetic profiles ($n_e$, $T_e$, $T_i$), equilibrium geometry, and NBI torque as inputs, and decomposes the momentum transport into diffusive, convective, and residual-stress contributions using physics-based scalings. The momentum diffusivity is linked to the ion heat diffusivity through an assumed Prandtl-number scaling~\cite{Strintzi_2008}, while intrinsic torque and pinch terms are evaluated from profile-gradient and shear-dependent residual-stress models. The resulting rotation profile is then obtained self-consistently from the momentum balance.

\begin{figure*}[t]
\centering
\includegraphics[width=0.85\textwidth]{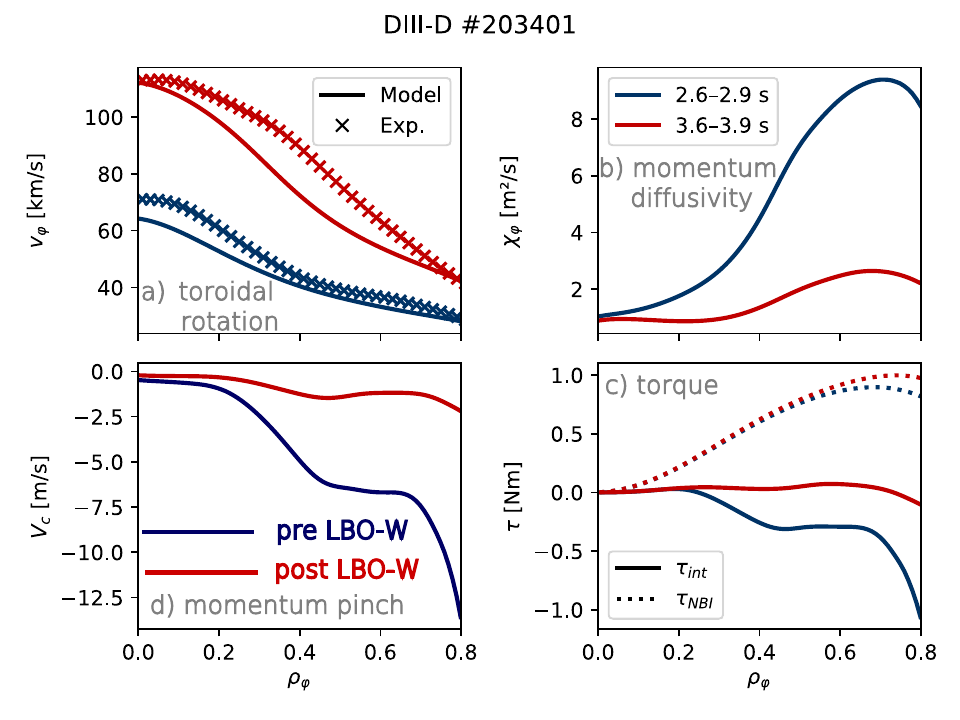}
\caption{Reconstructed toroidal rotation and momentum-transport coefficients before and after the LBO-W injection. 
(a) Ion toroidal rotation profiles from CER measurements (markers) together with the reconstructed rotation from the reduced rotation model~\cite{Zimmermann_2024}, shown for the phases before LBO-W ($t = 2.6$--$2.9\,\mathrm{s}$, blue) and with LBO-W ($t = 3.6$--$3.9\,\mathrm{s}$, red). 
(b) Ion momentum diffusivity $\chi_\phi$, showing substantially larger values before LBO-W, consistent with stronger underlying turbulence and larger ion heat diffusivity. 
(c) Intrinsic torque $\tau_{\mathrm{int}}$ (solid) together with the neutral-beam torque $\tau_{\mathrm{NBI}}$ (dashed). The NBI torque varies only weakly between phases, whereas the intrinsic torque changes significantly, indicating that the evolution of the rotation profile is primarily driven by changes in intrinsic momentum transport. 
(d) Momentum convective velocity $v_c$ (momentum pinch), where more negative values indicate stronger inward convection; the pre-LBO-W phase exhibits a stronger inward pinch.}
\label{fig:fig12-ben}
\end{figure*}

The model reproduces the observed rise in toroidal rotation during the W-cooling phase. This is illustrated in Fig.~\ref{fig:fig12-ben}(a), which shows the measurements as symbols and the modeling as solid lines, reproducing the correct trend. While the modeling relies on a boundary condition at the pedestal top, the rotation peaking in the core, well reproduced by the model, is stronger than the variation imposed at the boundary.

Shown in Fig.~\ref{fig:fig12-ben}(b), the ion momentum diffusivity $\chi_\phi$ is significantly larger before LBO-W, consistent with the stronger ITG/TEM turbulence and higher ion heat diffusivity in that phase. After W cooling, $\chi_\phi$ drops (in agreement with the reduction in $\chi_i$), allowing the rotation to rebuild even though the external NBI torque remains essentially unchanged. Shown in Fig.~\ref{fig:fig12-ben}(c), the intrinsic torque in this radial region relies mainly on turbulent residual stress and is taken into account by means of so-called profile-shearing effects~\cite{Camenen_2011}. Due to changes in the kinetic profile gradients (associated with the change in the turbulence regimes), it becomes less counter-rotating with LBO-W, supporting the recovery of toroidal rotation.

The momentum convective velocity $v_c$ (Fig.~\ref{fig:fig12-ben}(d)), relying on a scaling mimicking the Coriolis pinch~\cite{Peeters_2007}, is mainly coupled to the gradients of the electron density profiles. In the pre-LBO phase, it is more inward (more negative), whereas after W cooling it becomes less inward. This means that the change in $v_c$ is not due to an increase in the rotation drive; rather, without inward convection the difference in rotation between the two phases would be even larger. 

Physically, the strong electron cooling produced by W radiation counteracts the ongoing ECH heating, effectively removing a substantial fraction of the injected electron power. In this reduced-electron-heating state, the turbulence level falls, momentum diffusion decreases, and the rotation naturally rises. The effect observed herein is therefore very similar to the typical effects of counter-current intrinsic rotation in the presence of strong electron heating, discussed, for example, by McDermott \textit{et al.} for ASDEX Upgrade~\cite{McDermott_2011} or Grierson \textit{et al.} for DIII-D~\cite{Grierson_2019}.

In summary, the W-cooling phase reduces electron heating and suppresses TEM-like turbulence, while the increased $\ExB$ shear quenches the remaining ion-scale turbulence. Together, these effects weaken both ion and momentum diffusivities, naturally leading to an increase in core rotation. The explanation for the observed ion-temperature peaking lies in the fact that, after the W injection, edge ions become hotter than edge electrons and therefore act as an energy reservoir. With the core ion heat flux remaining essentially unchanged and both the ion diffusivity and pedestal ion temperature significantly reduced, the plasma must steepen the ion temperature gradient to exhaust the ion heat, resulting in the observed core $T_i$ peaking.

The changes in the plasma background state, now understood, must have important consequences for impurity transport, which are discussed in the following section.

\subsection{Impurity transport}
We discuss changes in impurity transport leveraging direct measurements of the carbon impurity density obtained with the CER system. The analysis focuses on carbon, which is directly measured, avoiding reliance on tungsten density reconstructions from SXR emissivity.

\begin{figure}[t]
\centering
\includegraphics[width=0.75\columnwidth]{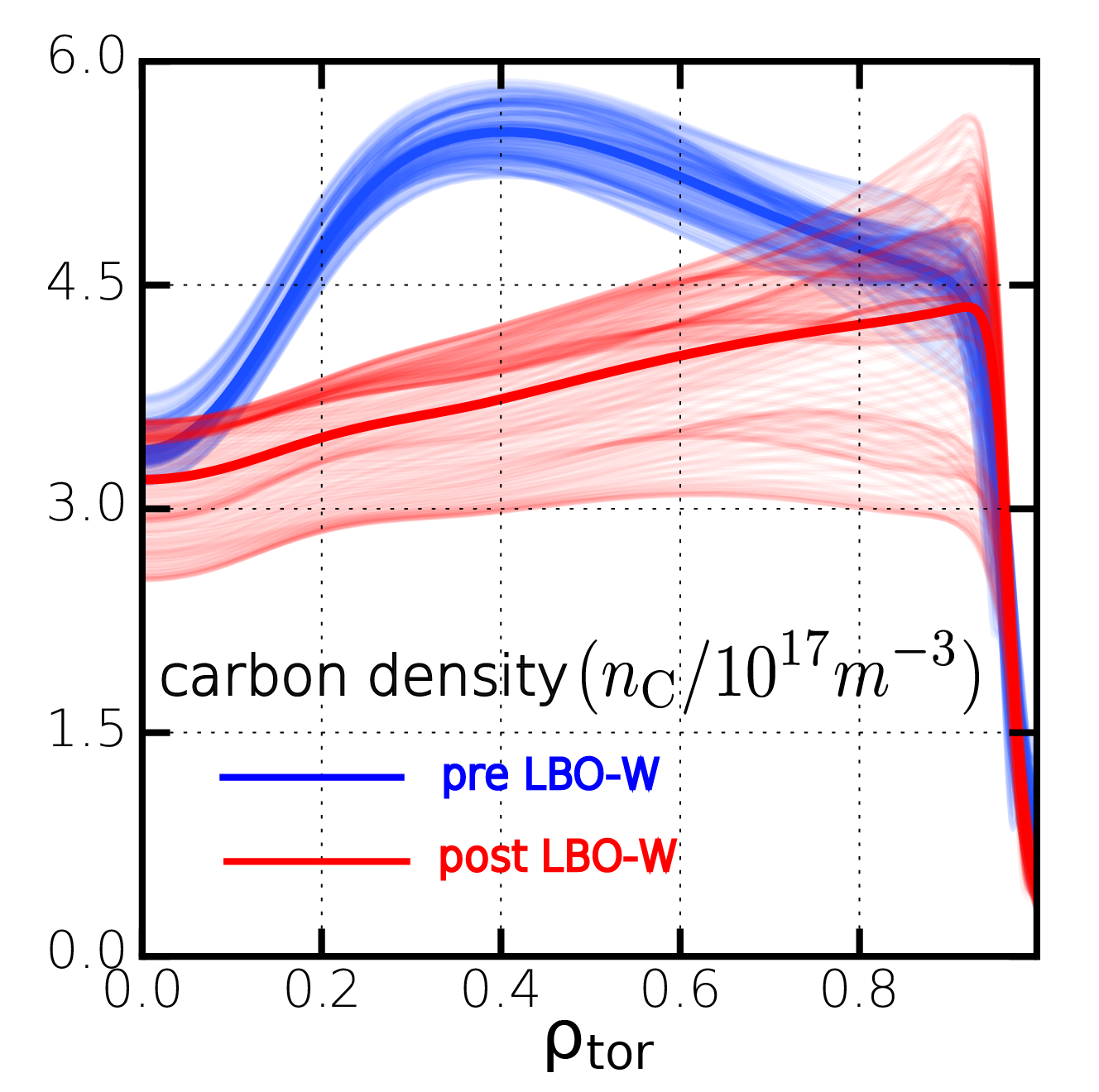}
\caption{Radial profiles of fully stripped carbon density measured by CER before (blue) and during (red) the W-cooling phase. The thin curves represent individual time slices within each phase. The carbon profile is strongly hollow on axis before the W injection, consistent with strong ECH heating. Carbon peaking increases during the W-cooling phase as the core pump-out--like hollowness disappears. These changes are accompanied by an overall reduction in carbon density in the plasma, providing evidence for modifications in impurity transport.}
\label{fig:fig13-carbon}
\end{figure}

Figure~\ref{fig:fig13-carbon} compares radial profiles of fully stripped carbon density before and during the W-cooling phase. Prior to W injection, the carbon profile is hollow in the plasma core during strong ECH. During the W-cooling phase, this hollow structure disappears and the profile becomes more centrally peaked. At the same time, the overall carbon density decreases across the radius. These observations indicate a modification of impurity transport during the W-cooling phase. 

The interpretation focuses on the central region, where turbulence is strongly reduced and neoclassical transport can become dominant, while impurity transport in the outer plasma remains primarily turbulence-driven. 

Neoclassical simulations performed with NEO are used to interpret the changes in the impurity profiles observed in Fig.~\ref{fig:fig13-carbon}. NEO computes neoclassical transport coefficients using the measured kinetic profiles, magnetic equilibrium, and plasma rotation, providing impurity diffusivity and convection for trace impurities. The shape of the impurity density profile is closely tied to the electron density and ion temperature gradients. A flatter electron density profile tends to reduce or even reverse the neoclassical inward convection, whereas a more peaked $n_e$ profile enhances it. In contrast, steep ion-temperature gradients weaken the inward convection through the known $T_i$-screening effect. This behavior follows directly from the neoclassical impurity flux expression 
\begin{equation}
\frac{R\Gamma_Z^{\mathrm{neo}}}{n_Z}
= D_Z^{\mathrm{neo}}
\left(
\frac{R}{L_{n_Z}}
- Z\frac{R}{L_n}
+ Z\frac{H_Z}{K_Z}\frac{R}{L_{T_i}}
\right)
\label{eq:neo_flux}
\end{equation}

where $D_Z^{\mathrm{neo}}$ is the neoclassical diffusion coefficient and the density and ion-temperature gradients drive inward and outward impurity transport, respectively. For convenience, the impurity flux can also be written in convection-diffusion form as  $\Gamma_Z^{\mathrm{neo}} = -D\nabla n_Z + Vn_Z$, where $V$ is the convection velocity (including inward pinch and the $T_i$-screening contributions). This yields the corresponding impurity peaking factor $V/D$.  

\begin{figure}[t]
\centering
\includegraphics[width=\columnwidth]{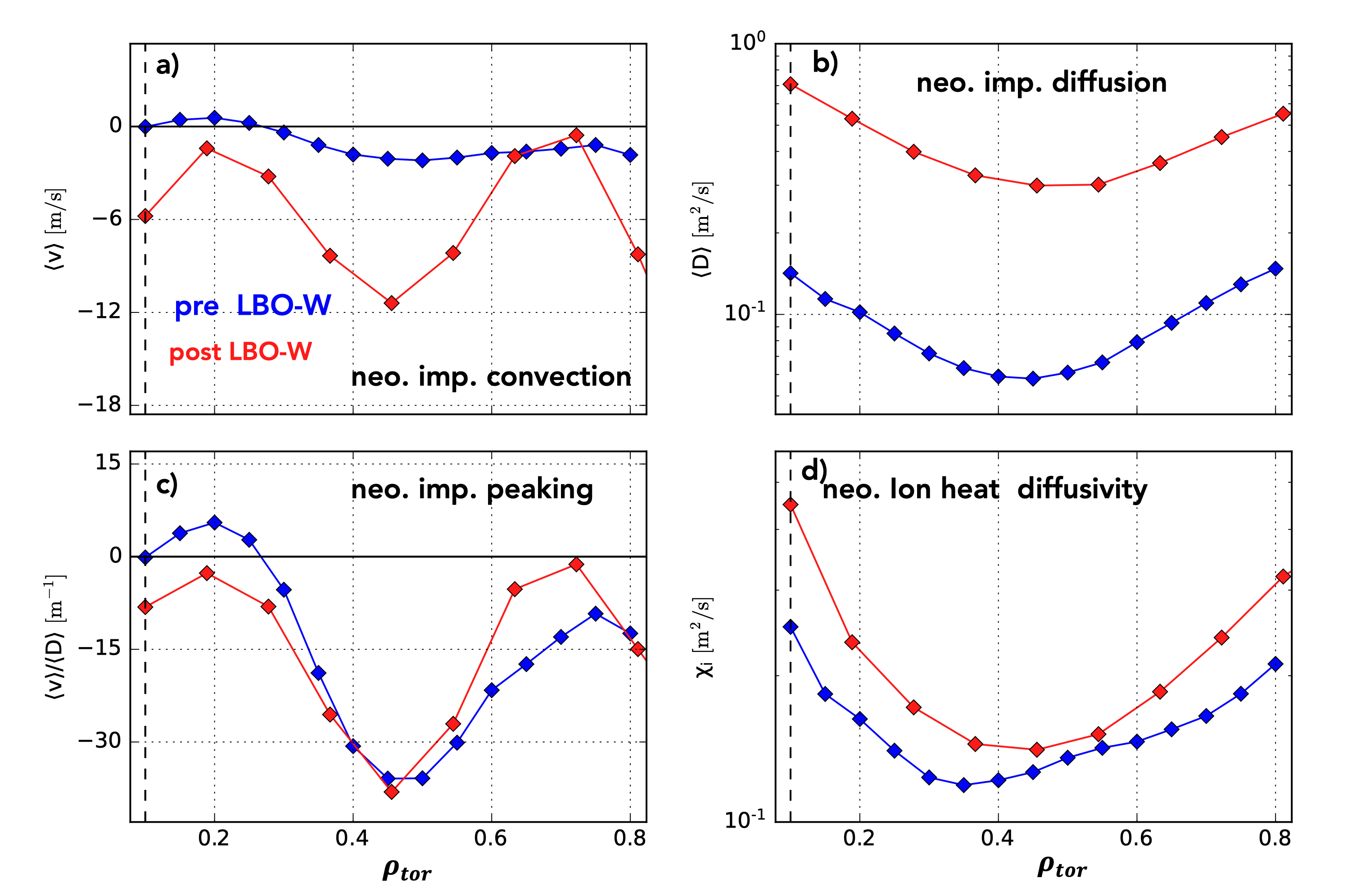}
\caption{Neoclassical impurity transport coefficients for carbon from NEO before and after the W injection. The panels show (top left) the neoclassical impurity convection $v$, (top right) impurity diffusivity $D$, (bottom left) the impurity peaking factor $v/D$, and (bottom right) the neoclassical ion heat diffusivity $\chi_i$, for the phases before LBO-W (blue) and during the W-cooling phase (red). The W-cooling phase exhibits a substantially stronger inward convection and larger diffusivity. NEO captures the positive (outward) value of $v/D$ in the core, responsible for the strong hollow impurity profile in the pre-cooling phase. These changes are consistent with the observed transition from a hollow to a more peaked carbon profile and reflect the increased importance of neoclassical transport as turbulence is reduced.}
\label{fig:fig14-neo}
\end{figure}

The NEO results in Fig.~\ref{fig:fig14-neo} show that during the W-cooling phase the neoclassical impurity convection $v$ becomes substantially more inward in the mid-radius region, while $D$ also increases. Increased plasma rotation is expected to play a major, possibly dominant role in this behavior, as both $D_{\mathrm{neo}}$ and the factor $H_Z/K_Z$ in Eq.~\ref{eq:neo_flux} are known to be strong functions of rotation ~\cite{Fajardo_2023}, supporting the enhanced $D_{\mathrm{neo}}$ and more inward pinch.

The resulting peaking factor $v/D$ becomes more negative, consistent with a stronger neoclassical inward pinch. These changes naturally explain the disappearance of the hollow carbon profile and the increase in core carbon peaking during W cooling as measured by CER. The corresponding neoclassical ion heat diffusivity $\chi_i^{(\mathrm{neo})}$ also increases in this phase, since the reduction in core turbulence makes the neoclassical channel relatively more important. A more detailed discussion of turbulent versus neoclassical contributions is provided in the next section.

While neoclassical transport explains the shape of the core impurity profile, it does not account for the overall reduction in total carbon content. The latter is likely governed by changes in pedestal particle transport, as the plasma enters a degraded-H-mode or intermittently L-mode-like state with poorer edge confinement. In this regime, the weakened particle confinement naturally leads to lower carbon density, even as core neoclassical peaking increases. A more detailed interpretation of this degraded pedestal and its impact on turbulence is provided in the discussion section.

\subsection{MHD activity}
A benign low-frequency MHD mode is present throughout the discharge, as is typical for DIII-D hybrid plasmas \cite{Turco_2023,Turco_2024a,Turco_2024b}. During the W-cooling phase its frequency increases by approximately a factor of two, while the overall mode character remains similar. We associate the increase in frequency with the rise in toroidal rotation during the W-cooling phase. The mode produces burst-like modulations in the core profiles and in the inferred impurity peaking. Only the pre-W-cooling phase is shown, where the lower mode frequency allows the modulation of the peaking factors to be more clearly resolved. Here the peaking factor is defined as the ratio of the central value to its value at $\rho_{\mathrm{tor}} = 0.4$, i.e.\ $f(0)/f(0.4)$, applied to the carbon density, ion temperature, and SXR emissivity.

\begin{figure*}[t]
\centering
\includegraphics[width=1\textwidth]{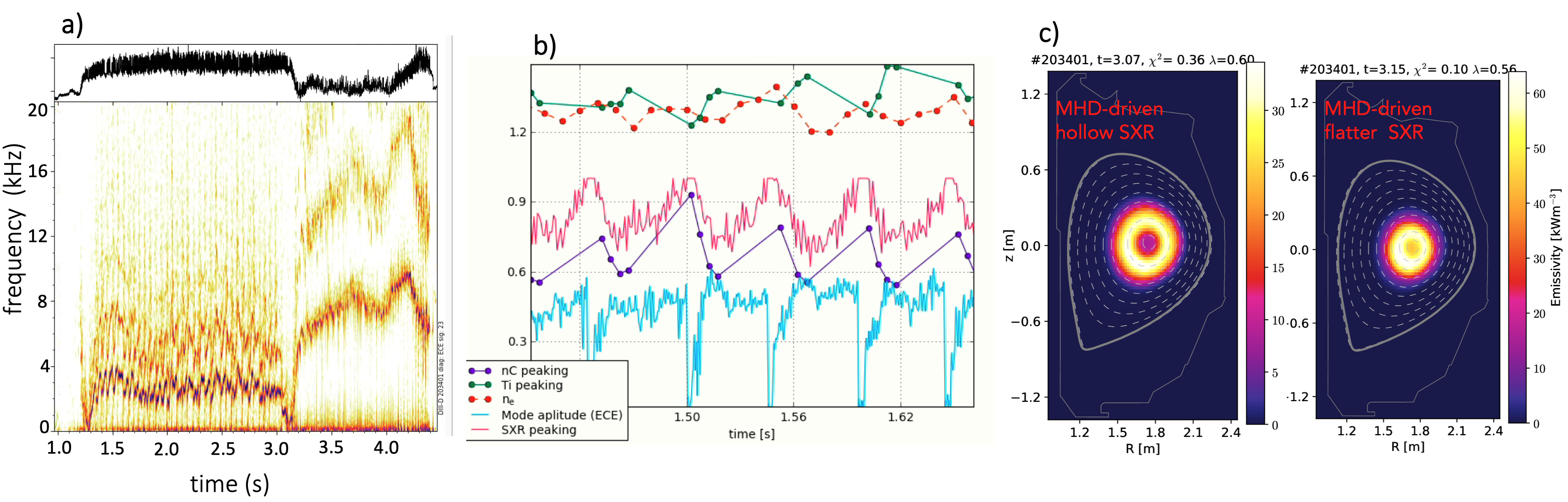}
\caption{(a) Spectrogram of magnetic fluctuations measured from ECE near the magnetic axis, showing the increase in frequency of the MHD mode following the W injection ($t > 3\,\mathrm{s}$). 
(b) Temporal evolution of electron density, carbon and ion-temperature peaking factors, together with the ECE mode amplitude and SXR peaking. The MHD activity oscillation modifies the SXR emissivity, correlated in time with density and temperature changes, producing cyclic hollowing and flattening of core quantities of the observed SXR radiation. 
(c) SXR tomographic reconstructions at two representative times, during flat and hollow emissivity phases induced by the MHD activity.}
\label{fig:fig15-mhd}
\end{figure*}

Fig.~\ref{fig:fig15-mhd}(b) shows that the carbon peaking factor, the ion-temperature peaking, the SXR emissivity peaking, and the ECE-derived mode amplitude correlate with the bursting of the MHD activity. The MHD cycle periodically redistributes heat and particles in the core, modulating both $T_i$ and $n_e$. These oscillations have direct neoclassical consequences:
\begin{itemize}
    \item periods of higher $T_i$ and flatter $n_e$ enhance the outward neoclassical convection through $T_i$-screening, reducing impurity peaking;
    \item periods of steeper $n_e$ and wealer $T_i$ restore stronger inward convection.
\end{itemize}

This interpretation is consistent with panel (b) of Fig.~\ref{fig:fig15-mhd}, where the variations of $T_i$-peaking and impurity peaking follow the MHD amplitude. Such redistribution is expected in hybrid plasmas, where benign core MHD interacts with transport without causing degradation of global confinement (e.g. no significant change in $H_98$, $\beta_N$, or disruption behavior).

Importantly, while MHD modulates the instantaneous impurity peaking, it does not account for the systematic change observed across the W-cooling transition. The underlying shift in turbulence regime and neoclassical transport, discussed in the preceding sections, dominates the time-averaged impurity behavior. The role of MHD here is therefore modulatory rather than causal, periodically adjusting the profiles around the mean set by turbulence and neoclassical transport.

\section{Discussions}
\label{sec5_discuss}
\subsection{H--L transition}
\begin{figure*}[t]
\centering
\includegraphics[width=0.85\textwidth]{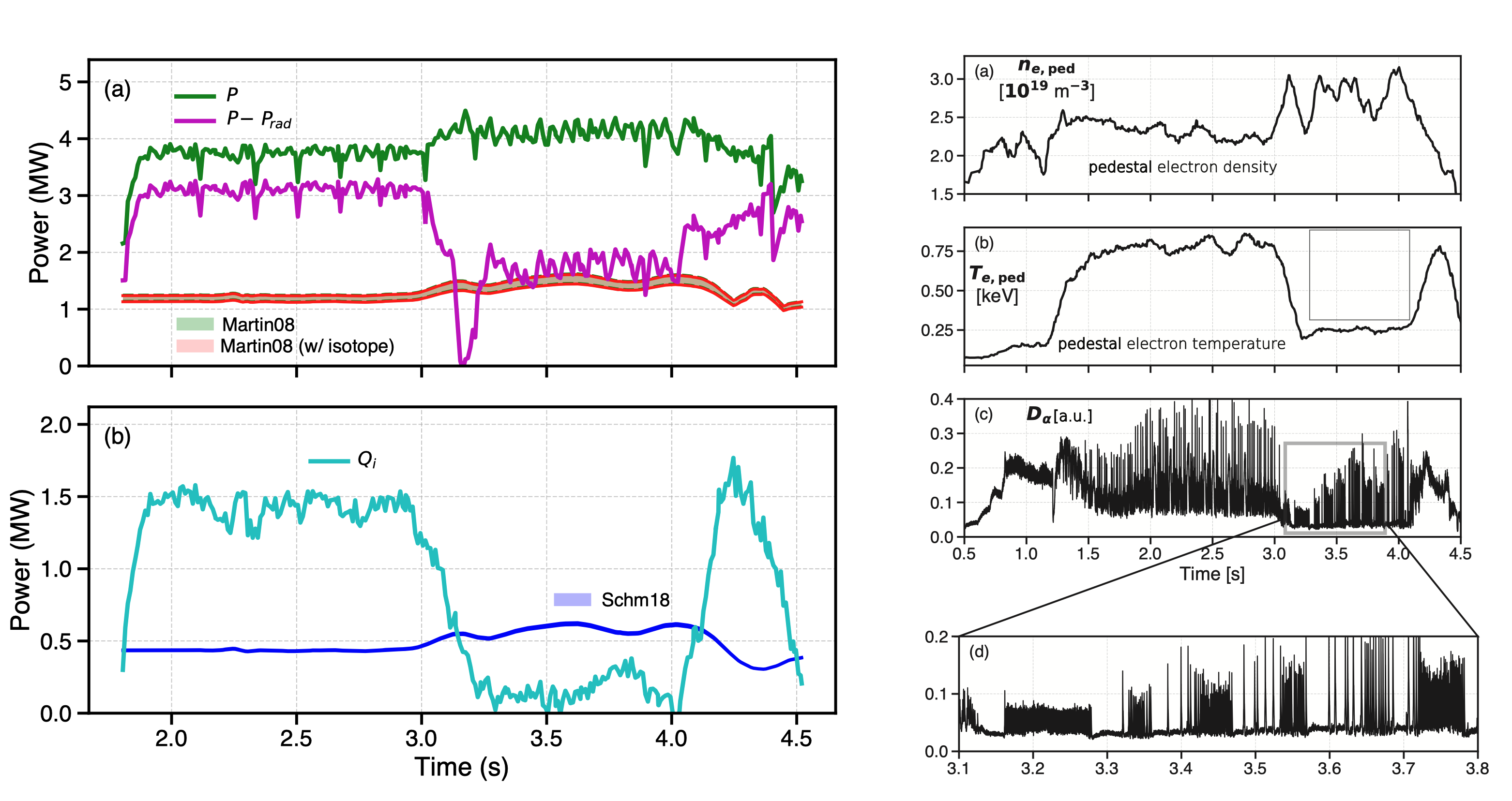}
\caption{Evolution of the plasma during the W-cooling phase and its approach to the H--L transition. Left column: (a) total input power, radiated power, and net heating power $P - P_{\mathrm{rad}}$, compared with the Martin threshold (with and without isotope correction); the net power approaches the H--L boundary during W cooling. (b) Ion heat flux compared with the Schmidmayr ion heat-flux threshold~\cite{Schmidtmayr_2018}, showing that the discharge approaches marginal H-mode conditions. Right column: pedestal electron density (c) and temperature (d). The pedestal temperature decreases while pedestal density increases during the W-cooling phase. Panels (e--f) show Filterscope $D_\alpha$ measurements and a zoom of the W-cooling phase, indicating intermittent edge activity. Together,thse observations indicate operation near the H--L boundary during W cooling, contributin to reduced particle confinement and decreased carbon content.}
\label{fig:fig16-lh}
\end{figure*}

A more detailed characterization of the L--H power threshold in this experiment is discussed in Ref.~\cite{Litaudon_2026}. The Martin scaling \cite{Martin_2008} and and the Schmidmayr ion heat-flux criterion \cite{Schmidtmayr_2018}, shown in Fig.~\ref{fig:fig16-lh}, indicate that during the W-cooling phase the discharge approaches the H--L boundary. The Schmidmayr ion heat-flux criterion corresponds to a threshold in the ion heat flux required to sustain H-mode confinement. The accompanying increase in intermittent $D_{\alpha}$ emission further supports the interpretation that the plasma operates near marginal H-mode conditions. As discussed previously in the context of impurity transport, the total carbon content decreases during the LBO-W injection, consistent with reduced particle confinement as the plasma approaches the H–L boundary under strong radiative cooling.

At these high radiated-power fractions and near-threshold $P_{\mathrm{sep}} = P_{\mathrm{inj}} - P_{\mathrm{rad}}$, strong tungsten-induced radiation may also lead to partial divertor detachment, as observed in high-$Z$ impurity-seeded DIII-D plasmas~\cite{Wang_2017}. However, assessment of detachment lies beyond the scope of the present work.

\subsection{Fast ions confinement}
\begin{figure*}[t]
\centering
\includegraphics[width=0.85\textwidth]{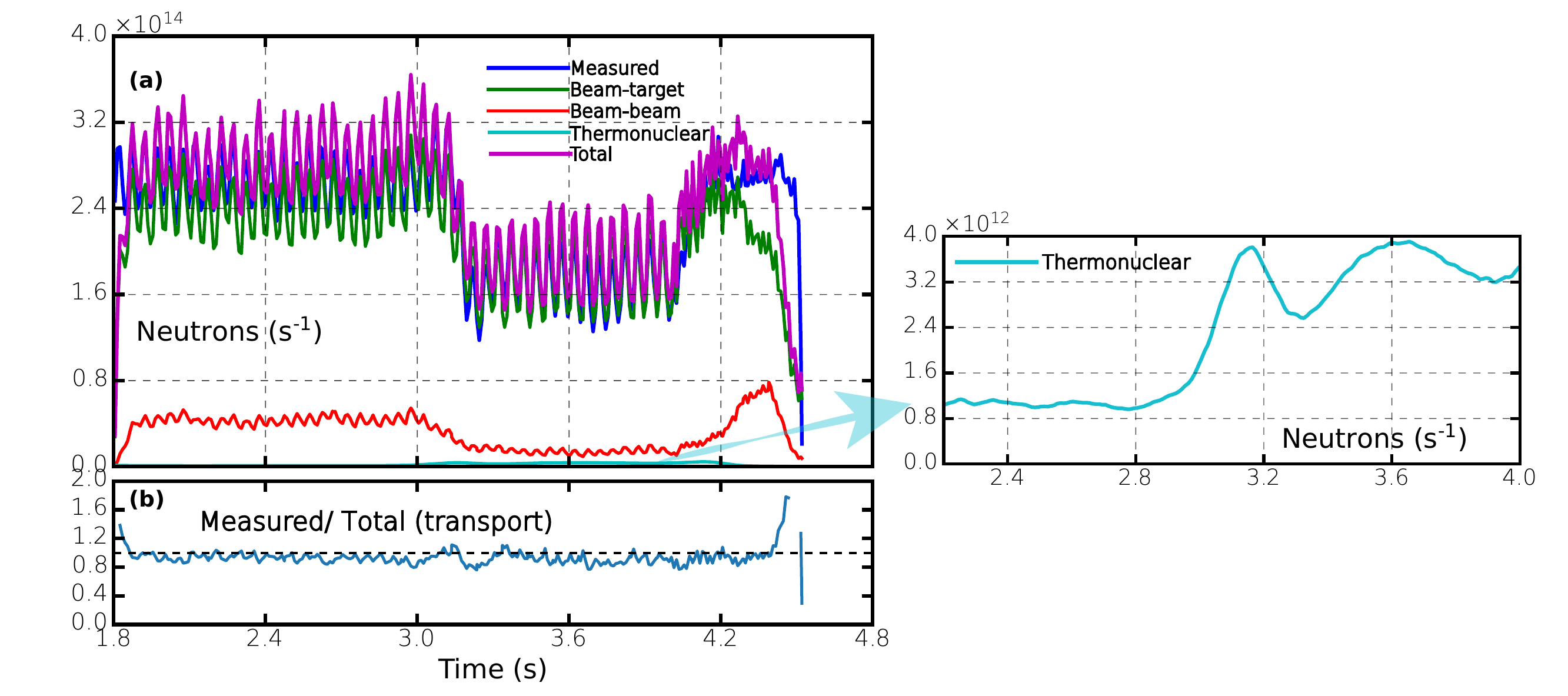}
\caption{Neutron emission and fast-ion behavior during the W-cooling phase.
Measured neutron rate compared with TRANSP/NUBEAM components: beam-target, beam-beam, thermonuclear, and total neutrons. The modeled neutron rate remains in good agreement with the measurement throughout the discharge, indicating no anomalous fast-ion losses during W cooling. During the W-cooling phase, the beam-driven neutron components decrease as increased density and reduced electron temperature shorten the classical slowing-down time, reducing the fast-ion population and pressure. In contrast, the thermonuclear neutron rate increases by approximately a factor of three due to the rise in core ion temperature. The ratio of measured to modeled neutron rate remains close to unity.}
\label{fig:fig17-fast-ion}
\end{figure*}

Fast-ion confinement was evaluated using the TRANSP NUBEAM \cite{Pankin_2004} analysis for this discharge, where NUBEAM is a Monte Carlo model for neutral beam deposition and fast-ion transport within TRANSP.  Despite the presence of benign MHD activity, the fast-ion population shows no evidence of anomalous deconfinement. The calculated classical total neutron rate from TRANSP agrees well with the measured neutron rate throughout the discharge, including the $3$--$4\,\mathrm{s}$ W-cooling phase. Because TRANSP uses purely classical slowing down and does not include anomalous fast-ion transport, this agreement indicates that neither the MHD activity nor the W injection induces additional fast-ion losses. During the W-cooling phase, the beam-target and beam-beam neutron components decrease (Fig.~\ref{fig:fig17-fast-ion}, indicating a reduction of the fast-ion population. 

This behavior can be explained by classical effects: the increase in electron density and the decrease in electron temperature shorten the energetic-ion slowing-down time, $\tau_s \propto T_e^{3/2}/n_e$, causing beam ions to thermalize more rapidly. As a result, the steady-state fast-ion population is reduced even though the injected NBI power remains constant. This reduced fast-ion density directly lowers the beam-driven neutron rate. Notably, the thermonuclear neutron rate increases during the W-cooling phase due to the rise in core ion temperature (i.e., the observed $T_i$ peaking and increased density), partially compensating for the reduced beam-target contribution. The combined effect is a net decrease in total neutron rate, dominated by the reduction of the fast-ion population while accompanied by an increase in thermal fusion reactivity.

Overall, the neutron emission and fast-ion pressure evolution during W cooling are fully consistent with classical slowing-down physics in a cooler, denser plasma, with no indication of anomalous fast-ion losses. We note that the ion population behavior naturally separates into two branches:
\begin{itemize}
    \item a thermal component, which increases due to $T_i$ peaking, with consequences of enhancing thermonuclear D-D reactions ($3 \times$ increase thermonuclear neutrons) and, 
    \item non-thermal fast-ion component, which decreases due to enhanced classical thermalization.
\end{itemize}

\subsection{The relative role of turbulent vs. neoclassical impurity transport }

The relative importance of turbulent and neoclassical impurity transport is assessed using the methodology introduced by Fajardo et al \cite{Fajardo_2024} for tungsten impurity, which compares the neoclassical diffusivity $D_{\mathrm{NC}}$ obtained from NEO with the turbulent diffusivity $D_{\mathrm{turb}}$ obtained from TGLF, using saturation rule 2. The key metric is the ratio $(Z D_{\mathrm{NC}} )/D_{\mathrm{turb}}$ , which indicates whether impurity transport is turbulence-dominated or neoclassical-dominated.

\begin{figure}[t]
\centering
\includegraphics[width=\columnwidth]{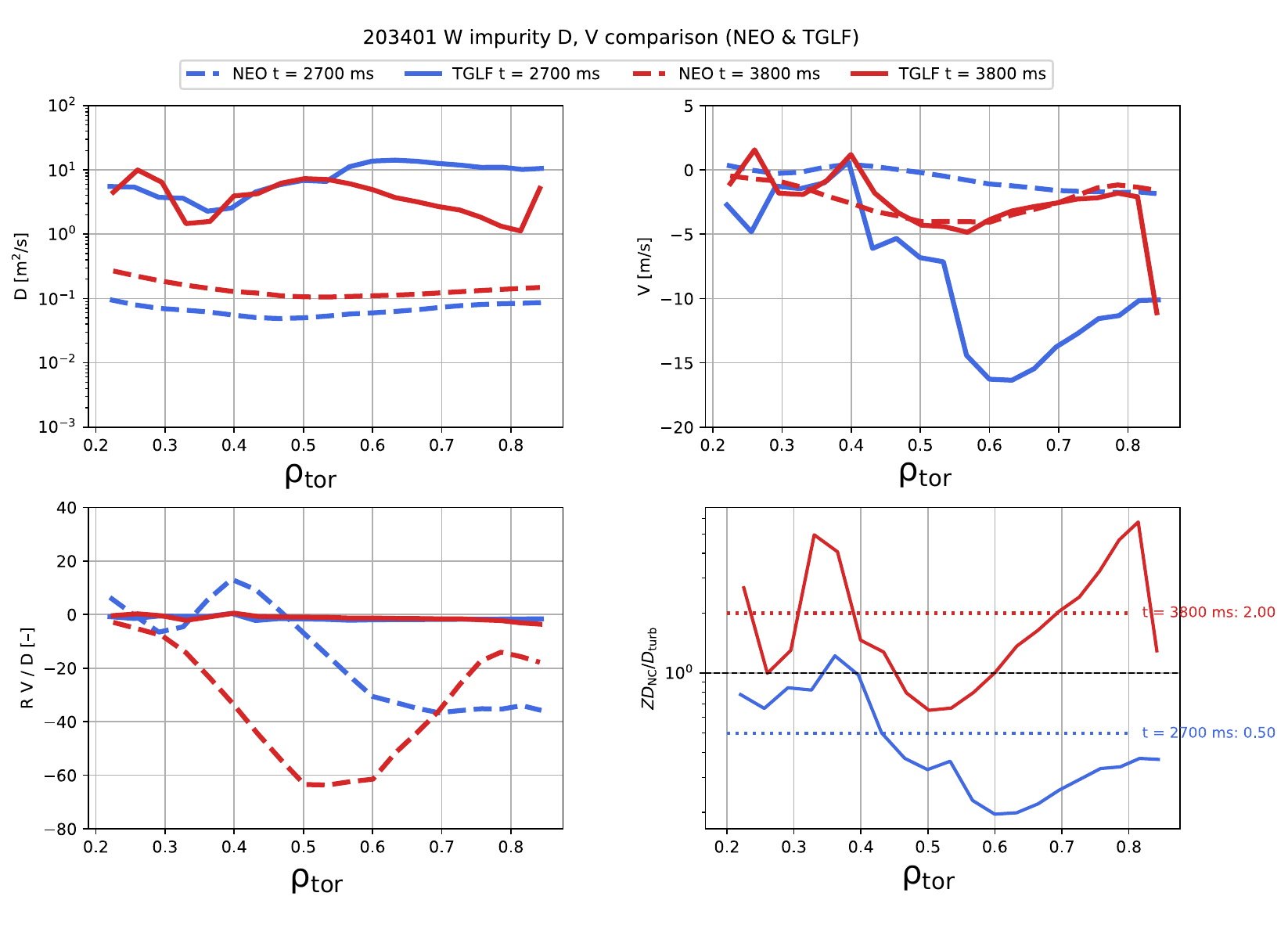}
\caption{Comparison of neoclassical (NEO) and turbulent (TGLF SAT2) impurity transport coefficients before ($2.7\,\mathrm{s}$, blue) and during the W-cooling phase ($3.8\,\mathrm{s}$, red). Top left: impurity diffusivity $D$, showing overall reduction in turbulent diffusivity during W cooling. Top right: neoclassical convective velocity $v$, which becomes more inward during W cooling due to changes in $n_e$ and $T_i$ gradients. Bottom left: impurity peaking factor $v/D$, indicating a stronger inward pinch in the cooled phase. Bottom right: ratio $(Z D_{\mathrm{NC}})/D_{\mathrm{turb}}$, following the methodology of Fajardo \textit{et al.}~\cite{Fajardo_2023}. Before W injection, turbulent transport dominates ($(Z D_{\mathrm{NC}})/D_{\mathrm{turb}} \sim 0.5$), whereas during W cooling the discharge transitions toward a regime with stronger neoclassical contribution ($(Z D_{\mathrm{NC}})/D_{\mathrm{turb}}>2$). The values are averaged across the profiles.} 
\label{fig:fig18-vd-neo-tglf}
\end{figure}

Before the W injection ($t = 2.7\,\mathrm{s}$), the discharge lies in a turbulence-dominated regime for tungsten transport, with $(Z D_{\mathrm{NC}})/D_{\mathrm{turb}} \sim 0.5$ on average across the radius (Fig.~\ref{fig:fig18-vd-neo-tglf}). 
In this phase, the turbulent diffusivity exceeds the neoclassical level by more than an order of magnitude. During the W-cooling phase ($t = 3.8\,\mathrm{s}$), strong radiative cooling reduces $T_e/T_i$, shifts the linear spectrum into ITG, and increases the $\ExB$ shearing rate. As shown by BES and TGLF, this suppresses ion-scale turbulence. Consequently, the turbulent diffusivity decreases, while the neoclassical convection becomes more inward due to enhanced plasma rotation. The increase in toroidal rotation corresponds to a moderate rise in the main-ion Mach number, previously reported to increase from $M_D \approx 0.16$ to $M_D \approx 0.32$ in the core, which further strengthens neoclassical impurity convection. 

\begin{figure}[t]
\centering
\includegraphics[width=\columnwidth]{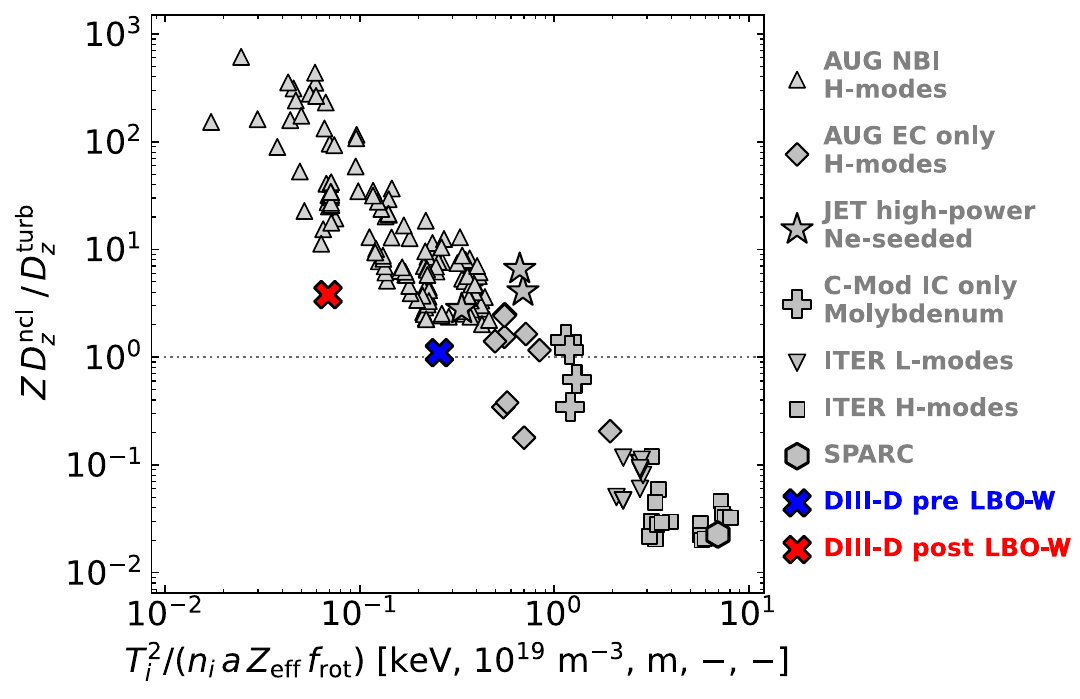}
\caption{DIII-D data points in the neoclassical-to-turbulent impurity transport parameter space following the framework of Fajardo \textit{et al.}. The ratio $Z D_{\mathrm{NC}}/D_{\mathrm{turb}}$ is shown as a function of $T_i^2/(n_i\, a\, Z_{\mathrm{eff}}\, f_{\mathrm{rot}})$ for multiple devices, including AUG, JET, C-Mod, ITER, and SPARC. The DIII-D data point shifts from the pre-LBO phase (blue) to the W-cooling phase (red), showing an increase in $Z D_{\mathrm{NC}}/D_{\mathrm{turb}}$ and thus a stronger neoclassical contribution to impurity transport. Under W cooling, the discharge moves away from ITER/SPARC-like conditions toward an ASDEX-Upgrade-like regime.}
\label{fig:fig19-fajardo}
\end{figure}

The net effect is a transition toward a regime with stronger neoclassical contributions to impurity transport, with $(Z D_{\mathrm{NC}})/D_{\mathrm{turb}} > 2$ during W cooling. This behavior aligns with the increased carbon peaking and the disappearance of the hollow core impurity profile observed with CER. Thus, consistent with the framework of Fajardo \textit{et al.}, the W-cooling phase drives the discharge from turbulence-dominated to neoclassical-dominated impurity transport mainly through the combined effects of reduced turbulence, modified background gradients, and enhanced rotation. We note that the DIII-D plasma considered here, initially closer to ITER and SPARC operating points in terms of the neoclassical-to-turbulent transport ratio, moves under W-cooling toward an ASDEX-Upgrade-like condition characterized by an enhanced role of neoclassical impurity convection, as illustrated in Fig.~\ref{fig:fig19-fajardo}. 

\subsection{Implications for WEST and future tungsten-walled devices}

The results presented here have direct relevance for WEST and other W-walled tokamaks, particularly with respect to core transport physics. Because WEST cannot directly measure plasma rotation, the DIII-D observations provide a controlled and well-diagnosed demonstration of how strong radiative cooling modifies the background state, stabilizes turbulence, increases toroidal rotation, and enhances neoclassical inward impurity convection.

Athough the present DIII-D plasmas are electron-dominated in terms of injected power, residual neutral beam injection provides direct ion heating that is absent in WEST, particularly influencing edge $T_i/T_e$.  In this respect, radiative-cooling experiments under pure ECRH heating, or with an increased fraction of electron heating in DIII-D, would offer an even closer comparison to WEST, especially for edge physics.

The results presented in this article also offer valuable preparation for a potential transition of DIII-D to a tungsten wall \cite{Abrams_2021}, clarifying how W-radiative cooling reshapes turbulence, rotation, and impurity behavior in a well diagnosed environment.

Altogether, these findings highlight that high-radiation tungsten regimes tend to counteract the turbulence-enhancing effects of ECH, shifting impurity transport toward neoclassical mechanisms, a trend of importance for ITER, SPARC, and future W-walled fusion devices, which must operate with low high-$Z$ core impurity accumulation. 

\section{Summary}
\label{sec6_conclusion}
This work presented the first detailed transport study of tungsten-induced radiative cooling in DIII-D hybrid-like plasmas operated under WEST-similarity constraints. Controlled LBO injection of tungsten (LBO-W) produced a transition to a high-radiation regime with $f_{\mathrm{rad}} > 0.5$, isolating the effects of tungsten radiative cooling on heat, momentum, and impurity transport. The high tungsten radiation is shown to reduce electron temperature, and thus the ratio $T_e/T_i$, which combined with an increase of $Z_{\mathrm{eff}}$, shifts the turbulence regime from mixed TEM/ITG toward a more weakly driven ITG-dominated state. The rise in toroidal rotation following W injection increased the $\ExB$ shearing rate, further suppressing ion-scale turbulence. Consistently, ion heat and momentum diffusivities decreased during the W-cooling phase, leading to a collapse of ion heat flux, a factor-of-two increase in toroidal rotation, and modest core $T_i$ peaking.

These transport changes strongly affected impurity behavior. Carbon measurements from CER showed that the hollow core impurity profile disappeared during the W-cooling phase and became more peaked, while the total carbon content decreased. NEO modeling indicated that reduced turbulence enhanced the relative importance of neoclassical transport, producing stronger inward convection in the plasma core. MHD activity modulated the instantaneous impurity peaking but did not explain the systematic change across the W-cooling transition, which was primarily set by the shift in turbulence and neoclassical transport.

The power-balance analysis showed that, despite the large increase in radiated power, radiative collapse did not occur. Instead, the plasma approached marginal H-mode conditions, with reduced particle confinement and signatures consistent with proximity to the H--L boundary. Neutron measurements remained consistent with classical TRANSP/NUBEAM predictions, indicating no anomalous fast-ion losses during W cooling.

Overall, tungsten-induced radiative cooling is shown to stabilize turbulence in DIII-D, reducing ion and momentum transport, and increasing the role of neoclassical impurity convection, even in plasmas with ECRH heating. These results are directly relevant for WEST and future tungsten-walled devices, including ITER and SPARC, where high-radiation operation must be reconciled with acceptable impurity accumulation and robust confinement.

\section{Acknowledgments}
The first author, A. T. B., gratefully acknowledges for C. Holland (University of California, San Diego), together with the Integrated Modeling group at MIT PSFC, for helpful input on turbulence stabilization. This material is based upon work supported by the U.S. Department of Energy, Office of Science, Office of Fusion Energy Sciences, using the DIII-D National Fusion Facility, a DOE Office of Science user facility, under Award(s) DE-FC02-04ER54698 and DE-SC0014264. 

\section{Disclaimer}
This report was prepared as an account of work sponsored by an agency of the United States Government. Neither the United States Government nor any agency thereof, nor any of their employees, makes any warranty, express or implied, or assumes any legal liability or responsibility for the accuracy, completeness, or usefulness of any information, apparatus, product, or process disclosed, or represents that its use would not infringe privately owned rights. Reference herein to any specific commercial product, process, or service by trade name, trademark, manufacturer, or otherwise does not necessarily constitute or imply its endorsement, recommendation, or favoring by the United States Government or any agency thereof. The views and opinions of authors expressed herein do not necessarily state or reflect those of the United States Government or any agency thereof.

\section*{References}
\bibliographystyle{unsrt}
\bibliography{biblio}

\end{document}